\documentclass{article}
\usepackage{epsfig}

\begin{document}
{\pagestyle{empty}
\parindent0pt\large\sf
LA-UR 97-4706\\~\\
Approved for public release;\\distribution is unlimited

\vfill

\begin{center}

\baselineskip3.0em

{\Huge\sf Two-lane traffic rules for cellular automata: A systematic
approach}

\vfill

Kai Nagel, Dietrich E Wolf, Peter Wagner, Patrice Simon

\end{center}

\vfill
\vfill

{\Huge\sf LOS ALAMOS}

{\sf N\,A\,T\,I\,O\,N\,A\,L~~L\,A\,B\,O\,R\,A\,T\,O\,R\,Y}

{\footnotesize\sf\openup-2pt
Los Alamos National Laboratory, an affirmative action/equal
opportunity employer, is operated by the University of California for
the U.S. Department of Energy under contract W-7405-ENG-36.  By
acceptance of this article, the publisher recognizes that the U.S.\
Government retains a non-exclusive, royalty-free license to publish or
reproduce the published form of this contribution, or to allow others
to do so, for U.S.\ Government purposes.  Los Alamos National
Laboratory requests that the publisher identify this article as work
performed under the auspices of the U.S. Department of Energy.  The
Los Alamos National Laboratory strongly supports academic freedom and
a researcher's right to publish; as an institution, however, the
Laboratory does not endorse the viewpoint of a publication or
guarantee its technical correctness. 
\vfill\eject
}
~\vfill\eject
}\setcounter{page}{1}
\title{%
Two-lane traffic rules for cellular automata: A systematic approach
}

\author{%
Kai Nagel,${}^{a,}$\protect\footnote{kai@lanl.gov}
Dietrich E.\ Wolf,${}^{b,}$\protect\footnote{d.wolf@uni-duisburg.de}
Peter Wagner,${}^{c,d,}$\protect\footnote{p.wagner@zpr.uni-koeln.de}
and
Patrice Simon${}^{a,}$\protect\footnote{simonp@tsasa.lanl.gov}\\
~\\
${}^a$ Los Alamos National Laboratory, TSA-DO/SA Mail Stop M997,\\
Los Alamos NM 87545, U.S.A.\\
${}^b$ Theoretische Physik, Gerhard--Mercator--Universit\"at,\\
47048 Duisburg, Germany\\
${}^c$ Zentrum f\"ur Paralleles Rechnen (ZPR), Universit\"at zu
K\"oln,\\
50923 K\"oln, Germany\\
${}^d$ Deutsches Zentrum f\"ur Luft- und Raumfahrt (DLR), Linder H\"ohe,\\
51170 K\"oln, Germany\\
~\\
DRAFT DRAFT \today\ DRAFT DRAFT
}

\maketitle

\begin{abstract}%
Microscopic modeling of multi-lane traffic is usually done by applying
heuristic lane changing rules, and often with unsatisfying results.
Recently, a cellular automaton model for two-lane traffic was able to
overcome some of these problems and to produce a correct density
inversion at densities somewhat below the maximum flow density.  In
this paper, we summarize different approaches to lane changing and
their results, and propose a general scheme, according to which
realistic lane changing rules can be developed.  We test this scheme
by applying it to several different lane changing rules, which, in
spite of their differences, generate similar and realistic results.
We thus conclude that, for producing realistic results, the logical
structure of the lane changing rules, as proposed here, is at least as
important as the microscopic details of the rules.
\end{abstract}

\section{Introduction}

Much progress has been made in understanding single lane traffic by
using simple models (e.g.~\cite{julich,tgf}).  Although one could
claim that these models also explain homogeneous multi-lane traffic,
they definitely fail when traffic on different lanes behaves
differently.  If one wants to investigate lane specific dynamics, one
has to address the question of how vehicles change from one lane to
the other. Here we propose an elementary scheme to develop such rules
and compare the simulation results of different realizations of this
scheme with empirical data from the German highway.

The preferred approach in science is to start from first principles and
then, using mathematics or simulation, to derive macroscopic
relationships. In sciences which involve human beings this is hopeless: the
gap between first principles and human behavior is too big.  One
alternative is to search heuristically for microscopically minimal
``plausible'' models which generate observed behavior on the macroscopic
level.  It is this approach that has often been used successfully when
physics methods have been applied in the area of socio-economic systems.
In this paper we want to go one step beyond that and look for systematic
logical structures in the rule sets for lane changing.

Accordingly, we start out from real world data (Sec.~2), followed by a
short review of traditional approaches to this problem in traffic
science (Sec.~3).  Sec.~4 outlines our approach.  In the following
three sections (Sec.~5 -- 7), we describe simulation results with
different rules.  Sec.~8 looks closer into the mechanism at flow
breakdown near maximum flow in the two-lane models.  Sec.~9 is a
discussion of our work, followed by a section showing how other
multi-lane models for cellular automata fit into our scheme (Sec.~10).
The paper concludes with a short summary.

\section{Real world measurements}
\begin{figure}
\caption{\label{Wiedemann}%
Traffic measurements in reality.  Left column: as function of flow;
right column: as function of density.  Top: flow; middle: velocity;
bottom: lane usage.  The units for density are vehicles per kilometer
per 2~lanes, and for flow they are vehicles per hour per 2~lanes.
Each data point corresponds to a 1~minute average.  Figure from
Wiedemann, see~\protect\cite{Wiedemann:2lane:data} for further information.
}
\end{figure}
As stated above, we are interested in {\em macroscopic\/} observations
of traffic flow quantities related to lane changing behavior.  A
typical such measurement can look like Fig.~\ref{Wiedemann}.  It
contains measurements of density (in vehicles/km/2\,lanes), flow (in
veh/h/2\,lanes), velocity (in km/h) and lane usage (in \%), all
averaged over one minute intervals.  The left column shows velocity
and lane usage as functions of flow; the right column shows flow,
velocity, and lane usage as functions of density. For theoretical
purposes, using flow as the control parameter has the disadvantage
that for the same flow value one has two different regimes---at high
density and at low density.  For example in the lane usage plot, one
cannot distinguish which data points belong to which regime.  We will
therefore concentrate on plots where density is the control parameter.

The top right plot shows the typical flow-density diagram.  Flow first
increases nearly linearly with density, until it reaches a maximum at
$\rho \approx 40$~vehicles/km/2\,lanes and $q \approx
3500$~vehicles/hour/2\,lanes.  {From} there, flow decreases with
increasing density, and the scatter of the values is much larger than
before. --- The currently best explanation for
this~\cite{Nagel:flow,Kerner:Konh:large:amplitude,Krauss:etc:metastable}
(but see also~\cite{Kerner:Rehborn:mea2,Kerner:Rehborn:prl}) is that,
for low densities, traffic is roughly laminar and jams are
short-lived.  In consequence, the addition of vehicles does not change
the average velocity much and flow is a linear function of density: $q
= \rho \, v$.  For high densities, traffic is an irregular composition
of jam waves, and laminar outflow traffic between jams.  Here, data
points are arbitrary averages over these regimes, leading to a much
larger variability in the measurements.

The plot of the velocity vs.\ density confirms this: There is an
abrupt drop in the average velocity at $\rho \approx
40$~veh/km/2\,lanes.  Yet, velocity is also not constant at lower
densities, leading indeed to a curvature of the flow-vs.-density curve
below the value $\rho \approx 40$~veh/km/2\,lanes, which can be
explained by the increasing influence of the slower vehicles in
multi-lane traffic.

The lane usage shows a peculiarity which is particularly strong in
Germany.  As should be expected, at very low densities all traffic is
on the right lane.\footnote{For countries such as Great Britain or
Australia, left and right have to be interchanged.} But with
increasing density, eventually more than half of the traffic is on the
left lane.  Only at densities above the maximum flow point, this
reverts to an equal distribution of densities between lanes.

Fig.~\ref{Wiedemann} does not show the flows of the individual lanes.
Ref.~\cite{Sparmann:2lane} contains such plots.  They show that the pointed
peak of the overall flow is caused by a pointed peak in the flow of the
left lane; flow on the right lane remains constant over a large density
range.

All this suggests the interpretation that the flow breakdown mechanism
on German autobahns is complicated, with flow breaking
down on the left lane first and thus not allowing the right lane to
reach its possible full capacity~\cite{Brilon:personal}.

\section{Traditional approaches}

Sparmann~\cite{Sparmann:2lane} discusses a lane changing
implementation for the microscopic Wiedemann-model~\cite{Wiedemann:model}.  Following Wiedemann's proposition, he distinguishes
between the wish to change lanes and the decision to change lanes.
For a lange change from right to left, these two parts
are:\begin{itemize}

\item
{\bf Wish} to change lanes if on any of the two lanes there is another
vehicle ahead and obstructing.

\item
{\bf Decision} to actually change lanes if there is enough space on
the other lane.

\end{itemize}
Conversely, for changing from left to right:\begin{itemize}

\item
{\bf Wish} to change lanes if on both lanes there is nobody ahead and
obstructing.

\item
{\bf Decision} to actually change lanes if there is enough space on
the other lane.

\end{itemize}
According to the philosophy of the Wiedemann-approach, ``obstructing''
is defined in terms of so-called psycho-physiological thresholds,
which depend mostly on speed difference and distance, and allow three
outcomes: no obstruction, light obstruction, severe obstruction.
Gipps~\cite{Gipps:2lane} reports a similar model.

The results are reported to be satisfying, yet unrealistic in at least
one respect: The density inversion between right and left lane near
maximum flow is not reproduced.

The Wiedemann-approach is a time-discrete formulation of a stochastic
differential equation and therefore continuous in space. Some recent
work in traffic has used a cellular automata approach, which is
coarse-grained discrete both in time and space.  Early lane changing
rules in the context of cellular automata models for traffic flow are
due to Cremer and co-workers~\cite{Cremer:Ludwig,Schuett}.
Following Sparmann, they implemented lane changes in the following
way: Lanes are changed to the left\begin{itemize}

\item
if a slower vehicle is less than $l_l$ cells ahead,

\item
and if a gap of size $\Delta x$ exists on the left lane;

\end{itemize}
lanes are changed to the right\begin{itemize}

\item
if, {\em on the right lane}, there is no slower vehicle less than $l_r$
cells ahead,

\item
and there is a gap of size $\Delta x$ on the right lane.

\end{itemize}
Again, they failed to reproduce the density inversion in the lane usage.


\section{Our approach}

Which contribution can Statistical Physics make in such a situation?  The
strength of Statistical Physics is to explain how microscopic relationships
{\em generate\/} macroscopic behavior.  Thus, the contribution of
Statistical Physics in traffic science (or in socio-economic systems in
general) will be to investigate which microscopic rules contribute to
certain aspects of macroscopic behavior and how.

Since current psychological knowledge does not allow to define beyond doubt
the set of microscopic rules involved in lane changing, we propose to
construct these rules according to certain symmetries inherent in the
problem.  As we will point out, these symmetries simplify considerably the
construction of consistent lane changing rules.  

Now, in spite of the absence of ``first principles'', it certainly
still makes sense to have a ``plausible'' starting point.  We thus
state here what we will use as the elementary laws, and later, how we
derive algorithmic rules from them.  Similar to
Ref.~\cite{Sparmann:2lane}, we propose that the basic ingredients are
security, legal constraints, and travel time minimization.  Security
requires to leave enough space between all vehicles.  The legal
constraints depend on the country.  Travel time minimization means
that one chooses the optimal lane under these constraints.

Let us start with security.  Security means that one leaves enough
space in front of and behind oneself.  As long as one stays on one
lane, this is ensured by single-lane driving rules, as e.g.\ given by
the rules in Refs.~\cite{Nagel:Prag,Nagel:Schreck}.  In the
context of changing a lane this means that there must be enough space
on the target lane.  Technically, one can say that there must be a gap
of size $gap_- + 1 + gap_+$.  The label $+$ ($-$) belongs to the gap
on the target lane in front of (behind) the vehicle that wants to
change lanes.  In the following we characterize the security criterion
by the boundaries $[-gap_-,gap_+]$ of the required gap on the target
lane relative to the current position of the vehicle considered for
changing lanes.

Different choices for both parameters are possible.  Throughout this
paper we use $gap_+ = v$ and $gap_- = v_{max}$ (i.e.~$[-v_{max},v]$),
where $v$ is the speed of the vehicle which changes lanes and
$v_{max}$ is the maximum velocity allowed in the cellular automaton.

Let us now go to legal constraints.  For example in Germany, lane
usage is regulated essentially by two laws: 1.~The right lane has to
be used by default, and 2.~passing has to be on the left.  In the
United States, the second law is considerably relaxed.  In this paper,
we will use ``Germany'' and ``United States'' as placeholders for two
somewhat extreme cases.  We expect that the behavior of many other
countries will be found somewhere in between.

Travel time optimization means that lane changes to the left are triggered
by a slow vehicle in the same lane ahead and when the target lane is more
attractive (because of optimization).  In this context, ``slow'' means a
velocity smaller or equal to the one of the car behind.  Here we give two
examples, first for changing to left:\begin{itemize}

\item[(a)] {\bf German criterion.}  In Germany passing is {\it not
allowed} on the right. Hence, if there is a slow vehicle on the {\it
left} lane, one has to change to the left, behind that slow
vehicle. Thus one changes to the left if there is a slow car ahead on
the same lane {\em or\/} on the left:
\[
v_r \le v \ .OR. \ v_l \le v \ .
\]

$v_r,v_l$ are taken within a certain distance one looks ahead, $d$,
which is a free parameter. If there is no vehicle within this
distance, the respective velocity is set to $\infty$.

\item[(b)] {\bf American criterion.}  By contrast, in America passing
on the right is not explicitely forbidden. The left lane is only more
attractive if the traffic there is faster than in one's own lane. Thus
one changes from the right to the left if there is a slower car ahead
in the same lane and if the next car in the left lane is faster than
the car ahead:
$$
v_r \le v \  .AND. \  v_r \le v_l \ .
$$

\end{itemize}

The easiest implementation of the law to use the right lane by default is
to make the criterion for changing back to the right lane the logical
negation of the criterion to change to the left lane; i.e.\ whenever the
reason to change to the left lane ceases to exist, one changes
back.\begin{itemize}

\item
This means for Germany that a change back to the right lane is tried as
soon as the velocities of the cars ahead in both lanes are sufficiently
large:
\[
v_r > v \ .AND. \ v_l > v \ .
\]

\item
In America, the rule would mean that one tries to change back if there
is a faster car than oneself (or no car at all) in the right lane, or
if traffic in the right lane is running faster than on the left lane:
\[
v_r > v \ .OR. \  v_r > v_l \ .
\]

\end{itemize}

In summary, a lane is changed if two criteria are
fulfilled:\begin{itemize}

\item
$\bullet$ Security criterion: $[-v_{max},v]$ are fulfilled.

\item
$\bullet$ Incentive criterion: Is there a good reason to change lanes?

\end{itemize}
The examples above illustrate that the wish to change from right to
left in general depends on both lanes. If the right lane is used by
default, the criterion to change from left back to right is that the
reason to change from right to left is no longer given, that is the
negation of the former criterion.

However, if the right lane is not used by default, it is natural to
consider symmetric incentive criteria: The return to the right lane
then depends on the same criterion as the transition to the left lane,
with ``left'' and ``right'' interchanged.  The simplest example, which
describes the {\em actual} American driving behavior fairly well,
involves only one lane in contrast to our ``American criterion''
above: One changes lanes only when a slow vehicle is
ahead:\begin{itemize}

\item
Criterion for change from right to left: $v_r \leq v$.

\item
Criterion for change from left to right: $v_l \leq v$.

\end{itemize}
This implies that vehicles stay on the left lane even when the right
lane is completely empty, and describes that American drivers often do
not use the rightmost lane in order to avoid the repeated disturbances
due to slow vehicles coming from on-ramps. In the words of symmetric
rules: When these drivers encounter {\em one} slow vehicle from on an
on-ramp, they switch to the left lane and stay there until they run
into a slower vehicle on that lane or until they want to get off the
freeway.  For that reason,
TRANSIMS~\cite{TRANSIMS,Nagel:etc:flow-char} in its current
microsimulation uses a totally symmetric lane-changing rule set.  See
Refs.~\cite{Nagatani:94:2lane,Rickert:diplom,Rickert:etc:2lane} for
symmetric lane changing rules.

Note that these considerations can easily be extended to multi-lane
traffic.  Also note that our paper only treats uni-directional traffic,
i.e.\ all vehicles are headed into the same direction.
Refs.~\cite{Schuett,Simon:Gutowitz} are examples for the treatment of
bi-directional traffic by cellular automata.

\section{Computer simulations of the basic velocity rules}

We now proceed to present computer simulations of the German rule-set
to illustrate the above principles.  Following Refs.~\cite{Latour,Rickert:diplom,Rickert:etc:2lane}, an update step of the whole system is
divided into two major substeps: (i)~lane changing, (ii)~forward
movement.

\subsection{Lane changing}

Lane changing here is implemented as a pure sideways movement.  One
should, though, better look at the overall result after the whole time
step is completed, and then lane changing vehicles usually will have
moved forwards, too.  Still, the algorithm is underestimating the time
vehicles usually need to change lanes: One CA iteration roughly
corresponds to one second; lane changes in reality need about
3~sec~\cite{Sparmann:2lane}.

More specifically, the lane changing algorithm is an implementation of the
following:

In {\em even\/} time steps, perform lane changes from right to
left.\footnote{%
We separate changes from left to right and changes from right to left
in anticipation of three lane traffic.  In three lane traffic, in a
simultaneous update it is possible that a vehicle from the left lane
and a vehicle from the right lane want to go to the same cell in the
middle lane.  {From} a conceptual viewpoint of simulation, this may be
called a scheduling conflict.  Such conflicts can be resolved by,
e.g., different update schedulings (such as here)~\cite{Barrett:personal,Barrett:theosim1}.
} All vehicles on the right lane for which the Incentive Criterion
($v_r \le v \ .OR.\ v_l \le v$) and the Security Criterion
($[-v_{max},v]$) are fulfilled are simultaneously moved to the left.

In {\em odd\/} time steps, perform lane changes from left to right.
All vehicles on the left lane for which the Incentive Criterion ($v_r
> v \ .AND.\ v_l > v$) and the Security Criterion ($[-v_{max}, v]$)
are fulfilled are simultaneously moved to the right.

The number of sites one looks ahead for the Incentive Criterion plays
a critical role.  Quite obviously, if one looks far ahead, one has a
tendency to go to the left lane already far away from an obstructing
vehicle, thus leading to a strong density inversion at low densities.
Thus, this parameter can be used to adjust the density inversion. ---
The results described below were obtained with a lookahead of
16~sites, that is, if no vehicle was detected in that range on that
lane, the corresponding velocity $v_r$ or $v_l$ was set to $\infty$.

\subsection{Forward movement}

The vehicle movement rules~(ii) are taken as the single lane rules
from Nagel and Schreckenberg~\cite{Nagel:Prag,Nagel:Schreck}
which are by now fairly well understood~\cite{Nagel:flow,Sasvari:Kertesz,Eisenblaetter:etc}.

For completeness, we mention the single lane rules here.  They
are\begin{itemize}

\item
IF ( $v < v_{max}$ ) THEN $v := v + 1$ \ \ (accelerate if you can)

\item
IF ( $v > gap$ ) THEN $v := gap$ \ \ (slow down if you must)

\item
IF ( $v \ge 1$ ) THEN WITH PROBABILITY $p$ DO $v := v-1$ \ \
(sometimes be not as fast as you can for no reason).

\end{itemize}
These rules for forward movement will be used throughout the paper.  ,
with $p$ equal to $0.25$.  All simulations are performed in a circle
of length $L=10\,000$.  The maximum velocity is $v_{max}=5$.

\subsection{Results}

\begin{figure}

\centerline{\hfill \epsfxsize0.49\hsize\epsfbox{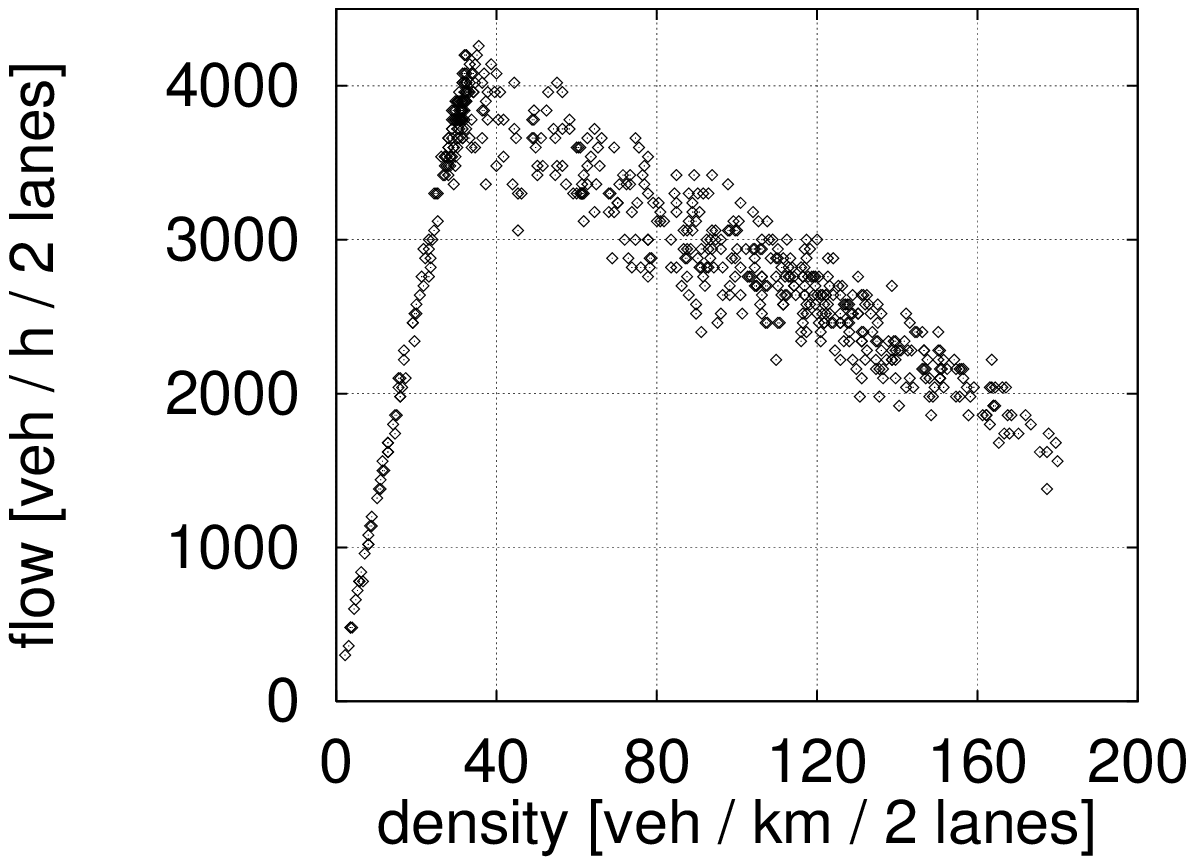}}
\centerline{\epsfxsize0.49\hsize\epsfbox{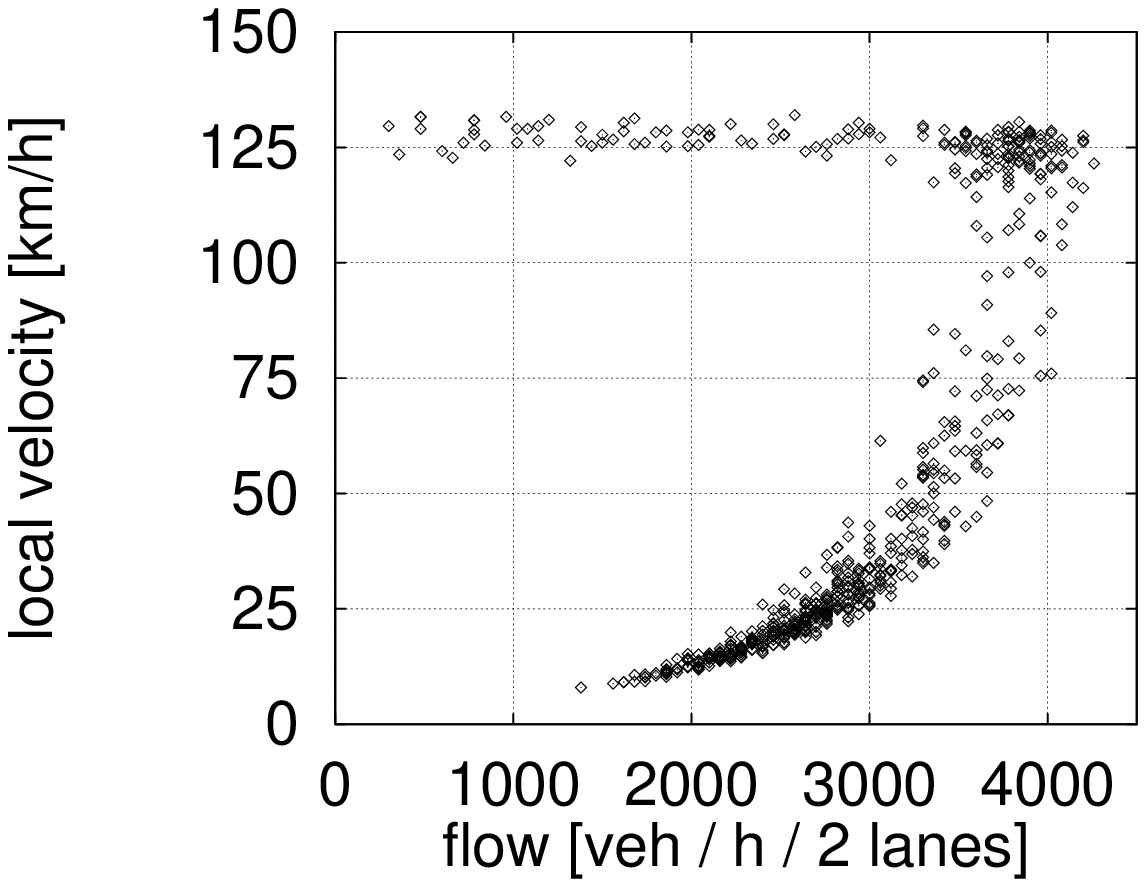} \hfill \epsfxsize0.49\hsize\epsfbox{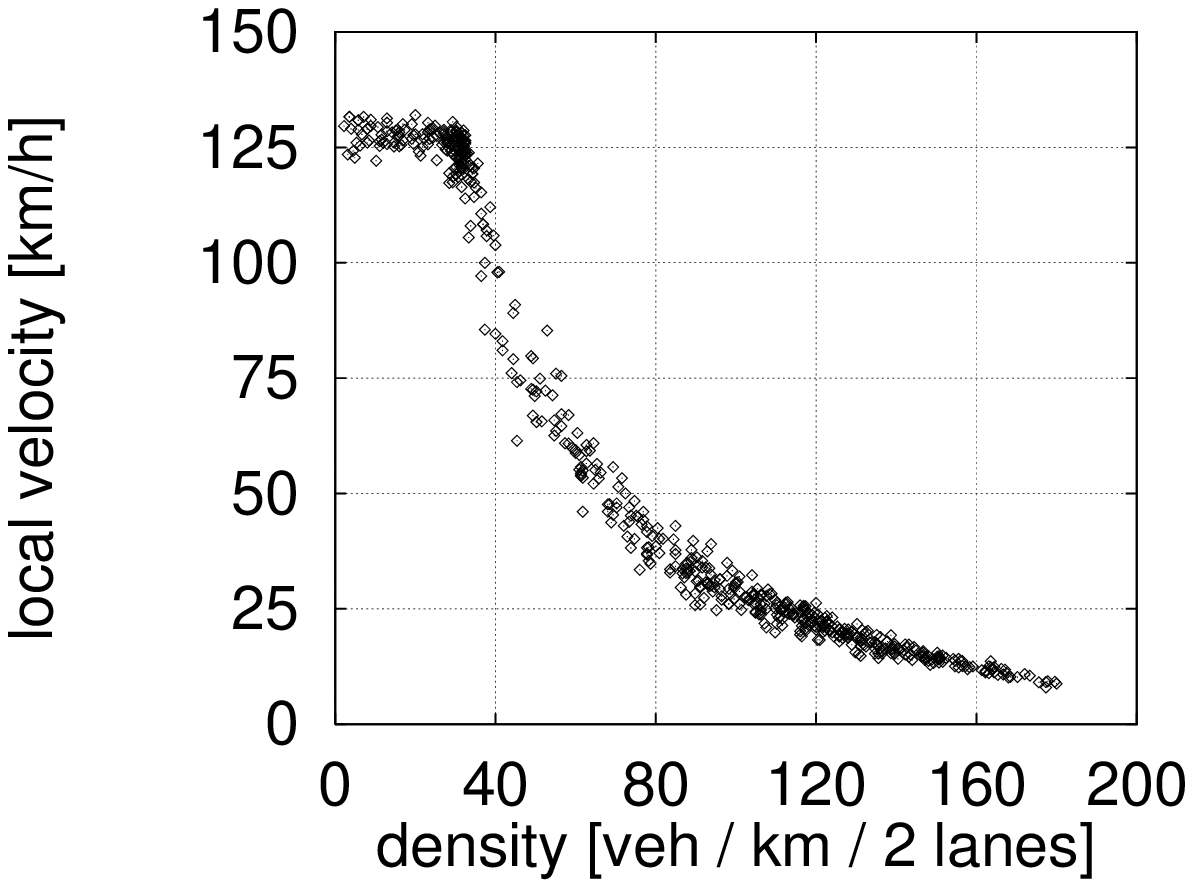}}
\centerline{\epsfxsize0.49\hsize\epsfbox{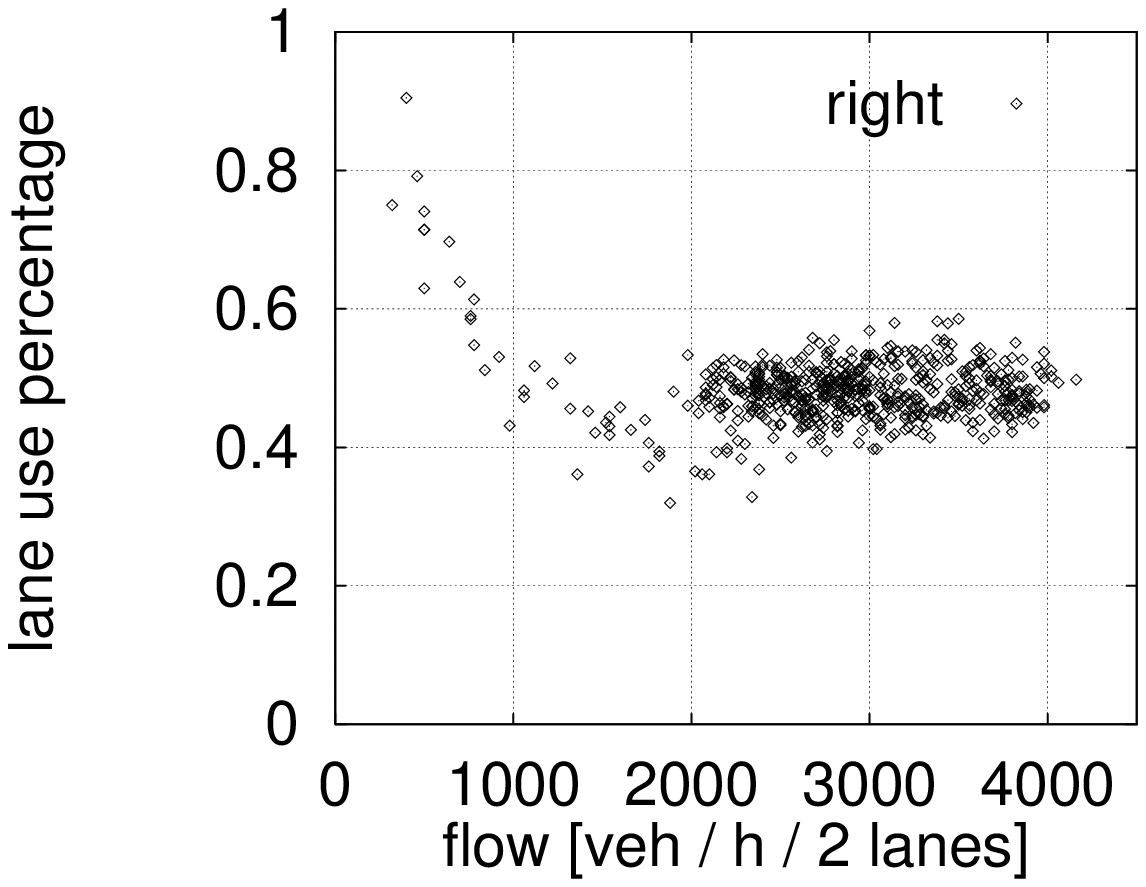} \hfill \epsfxsize0.49\hsize\epsfbox{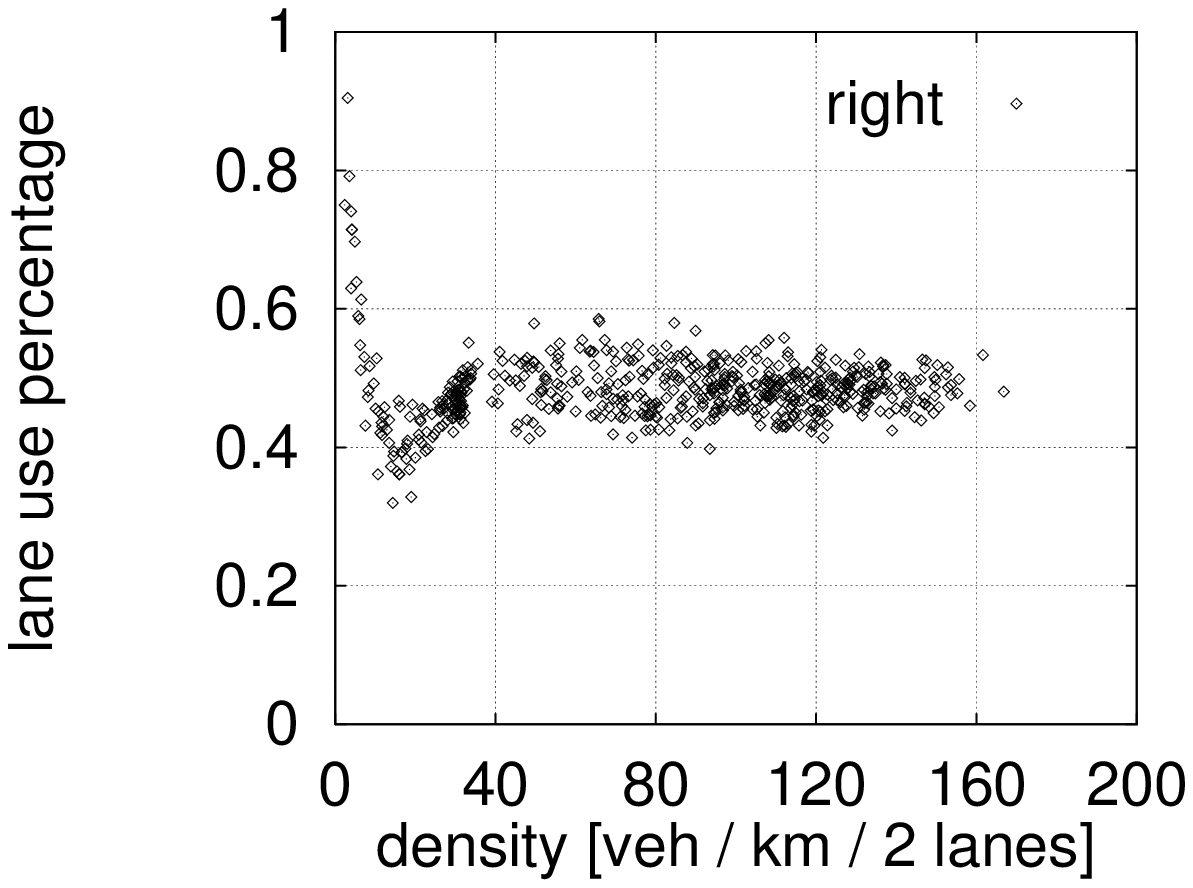}}

\caption{\label{basic:v}%
Simulation results for basic version of the velocity-based lane changing
rules.  Same type of plots as in Fig.~\protect\ref{Wiedemann}.  Each data
point is a one minute average, except for lane usage, where each data point
is a three minute average.
}
\end{figure}

As shown in Fig.~\ref{basic:v}, these rules generate reasonable
relations between flow, density, and velocity.  More importantly, they
generate the density inversion below maximum flow which is a so
important aspect of the dynamics on German freeways.  Note that, maybe
contrary to intuition, it is not necessary to have slow vehicles in
the simulations in order to obtain the density inversion.

\section{Computer simulations of gap-rules}

For comparison, we also simulated a version of Wagner's
``gap-rules''~\cite{Wagner:julich,Wagner:etc:2lane}, which is
adapted to our classification scheme above.  The reason to change to
the left then becomes
\[
gap_r < v_{max} \ .OR. \ gap_l < v_{max} \ ,
\]
i.e.\ one has a reason to change to the left when there is not enough
space ahead either on the right or on the left lane.

As stated above, as reason to change to the right we take the
negation, although we allow for some ``slack'' $\Delta$:
\[
gap_r \ge v_{max} + \Delta \ .AND.\
gap_l \ge v_{max} + \Delta \ ,
\]
i.e.\ one changes from left to right if on both lanes there is
enough space ahead.

The ``slack'' parameter $\Delta$ has been introduced in
Ref.~\cite{Wagner:julich}. The larger it is the less inclined is the
driver to change back to the right lane, and hence the more pronounced
is the lane inversion. In this sense the parameter $Delta$ plays a
similar r\^ole in these gap-rules as the look-ahead distance in the
basic velocity rules discussed before.  We will use $\Delta=9$, the
same value as in Ref.~\cite{Wagner:etc:2lane}.

\begin{figure}

\centerline{\epsfxsize0.49\hsize\epsfbox{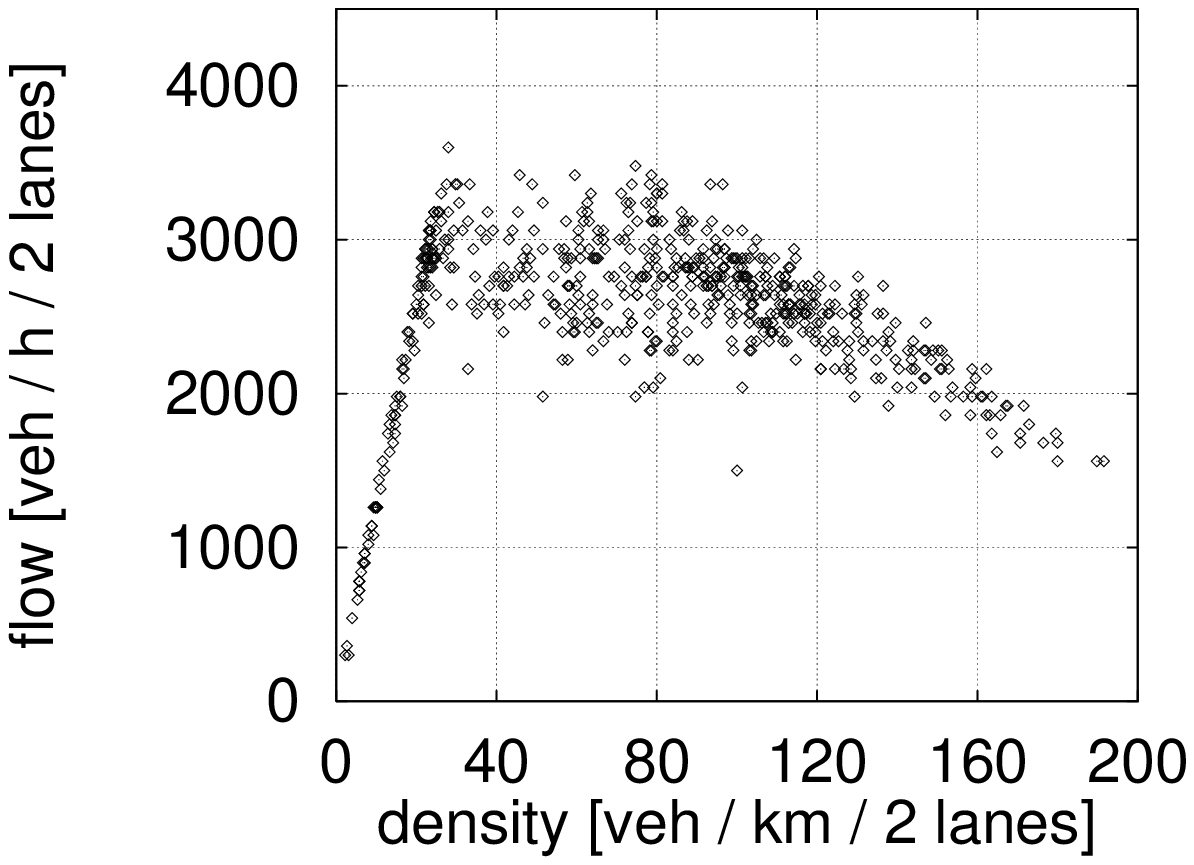}} 
\centerline{\epsfxsize0.49\hsize\epsfbox{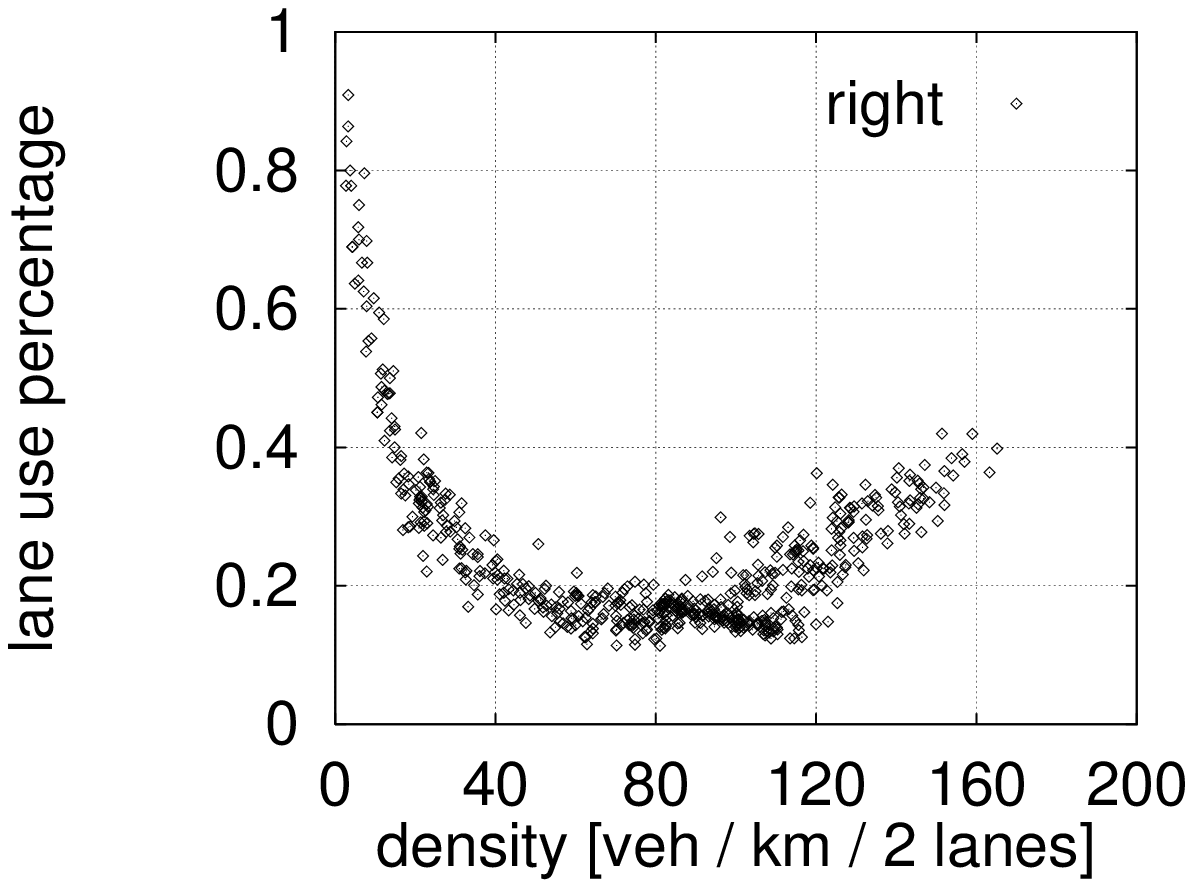}}
\caption{\label{gap}%
Simulation results for gap-based lane changing rules.  {\em Top:}
Flow vs.\ density.  {\em Bottom:} Lane usage vs.\ density.
}
\end{figure}

Fig.~\ref{gap} shows results of simulations with these rules.  One
immediately notes that these rules both qualitatively and quantitatively
generate the correct density inversion at maximum flow, i.e.\ at $\rho
\approx 38$~veh/km/2\,lanes; but from there on with further increasing
density the density inversion increases further, contrary to reality.
Ref.~\cite{Wagner:etc:2lane} uses rules which (i)~prohibit passing on the
right and (ii)~symmetrize traffic at very high densities; as a result, lane
usage becomes much more symmetric above the density of maximum flow.

\section{Extensions for reality}
After having shown that both velocity-based and gap-based lane
changing rules, based on the introduced logical scheme, can generate
the density inversion effect, we now proceed to include more realism
to bring the result closer to Wiedemann's data (Fig.~\ref{Wiedemann}).

\subsection{Slack}
\begin{figure}
\centerline{\epsfxsize0.49\hsize\epsfbox{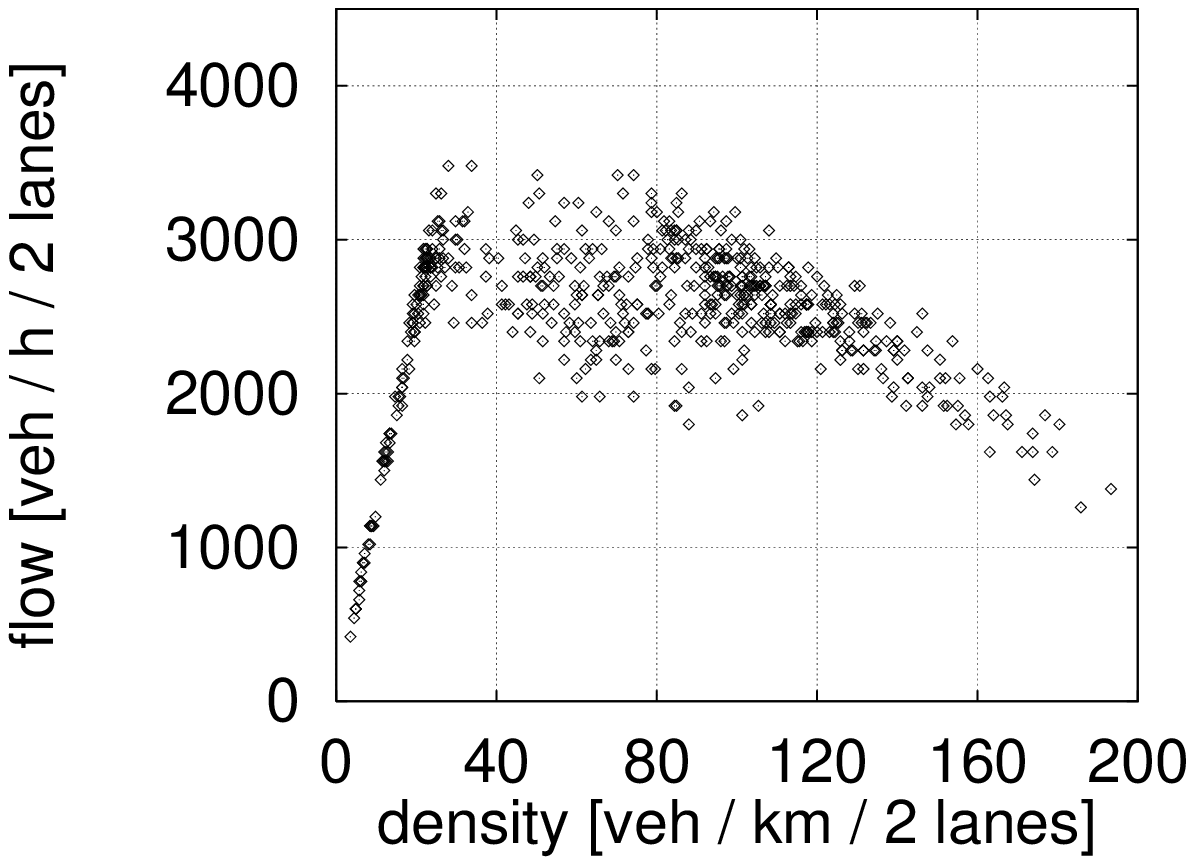}} 
\centerline{\epsfxsize0.49\hsize\epsfbox{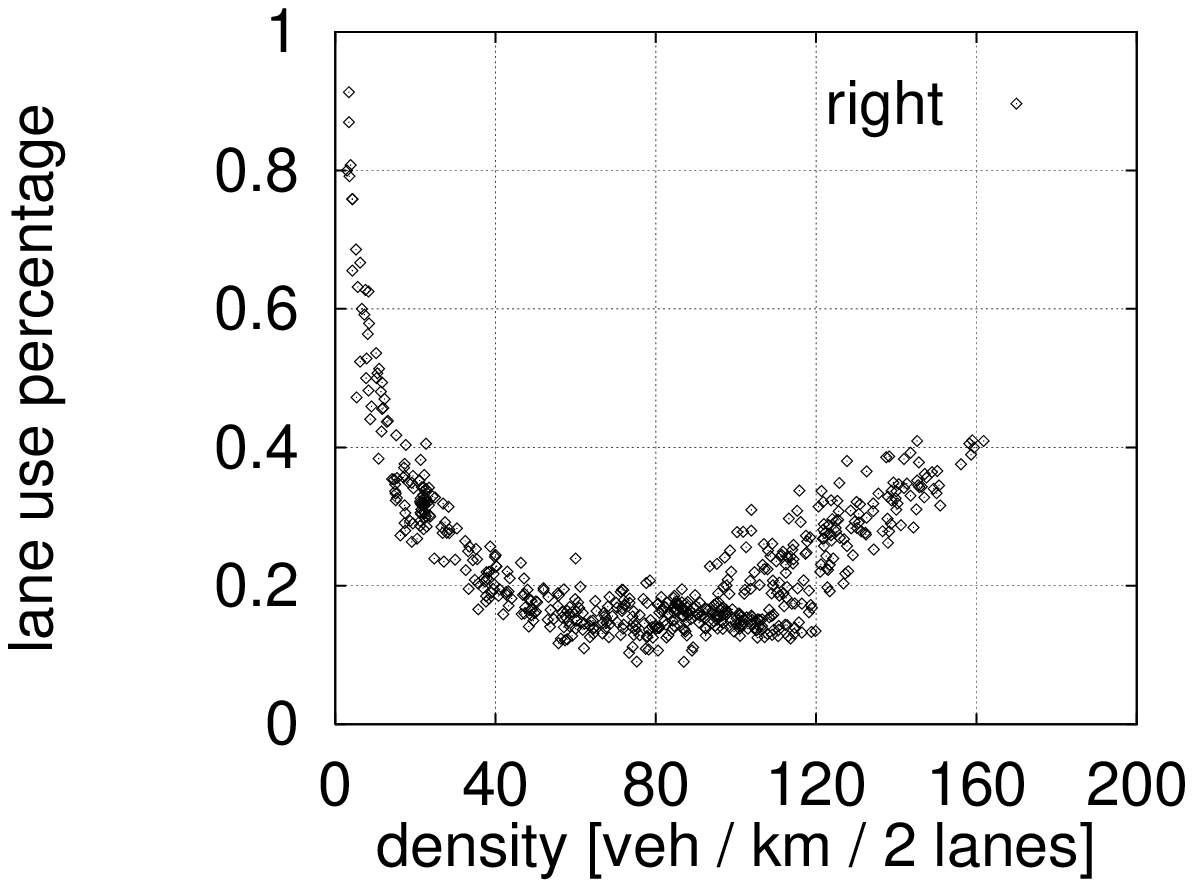}}
\caption{\label{v.w.slack}%
Simulation results for velocity-based lane changing rules with slack
(i.e.\ there is some ``slack'' in the incentive criterion for changing
to the right compared to the one for changing to the left).  {\em
Top:} Flow vs.\ density.  {\em Bottom:} Lane usage vs.\ density.
}
\end{figure}

With the basic velocity-based rules, one can adjust the density
inversion to the correct lane use percentage, but the maximum
inversion is reached at too low densities (at approx.\
16~veh/km/2\,lanes compared to approx.\ 28~veh/km/2\,lanes in
reality).  One possibility to improve this is to introduce some slack
$\Delta=3$ into the rules similar to the slack in the gap-based rules,
i.e.\ vehicles change to the left according to the same rules as
before, but the Incentive Criterion for changing back is not the
inversion of this. Instead, it now reads
\[
v_r > v + \Delta \ .AND. \ v_l > v + \Delta \ .
\]
Since these rules tend to produce a stronger density inversion than
before, we reduced the look ahead value to~7 to obtain realistic lane
usage values.  Results are shown in Fig.~\ref{v.w.slack}.

\subsection{Slack plus symmetry at high densities/low velocities}
\begin{figure}
\centerline{\epsfxsize0.49\hsize\epsfbox{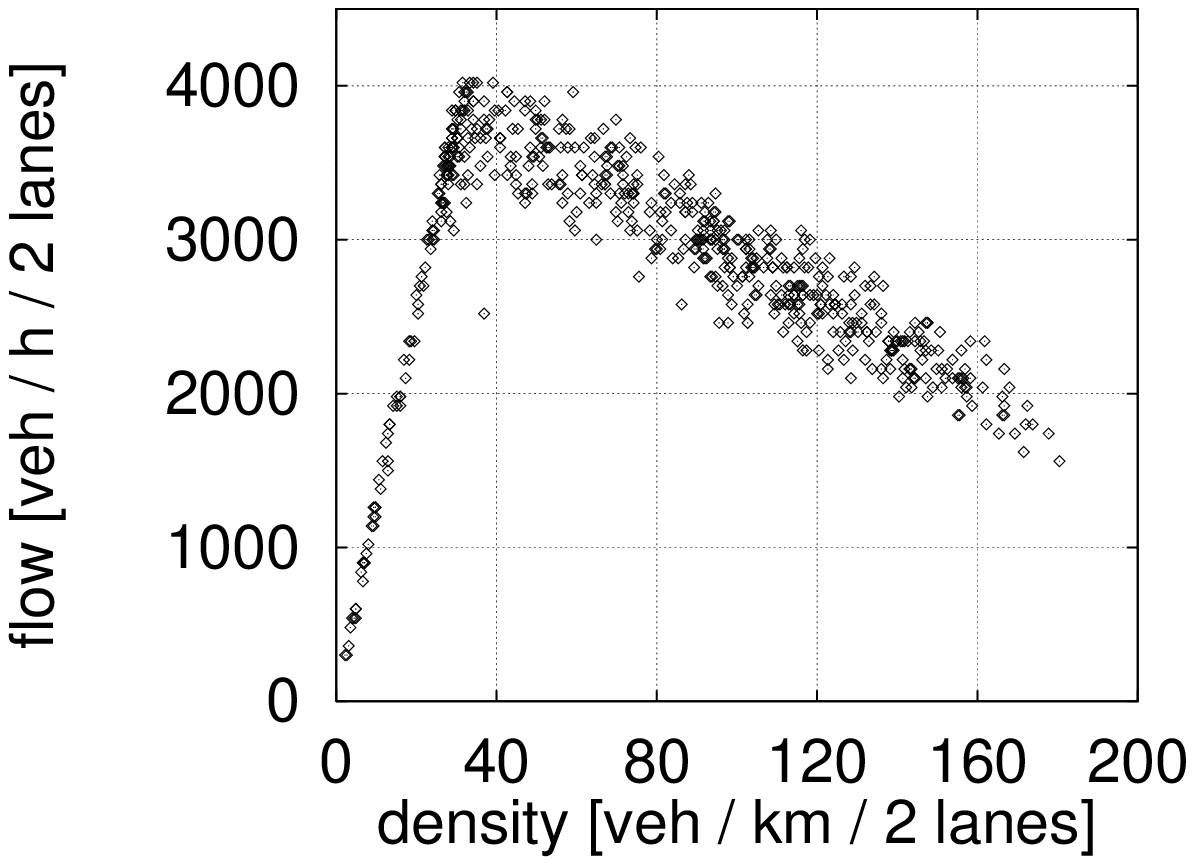}} 
\centerline{\epsfxsize0.49\hsize\epsfbox{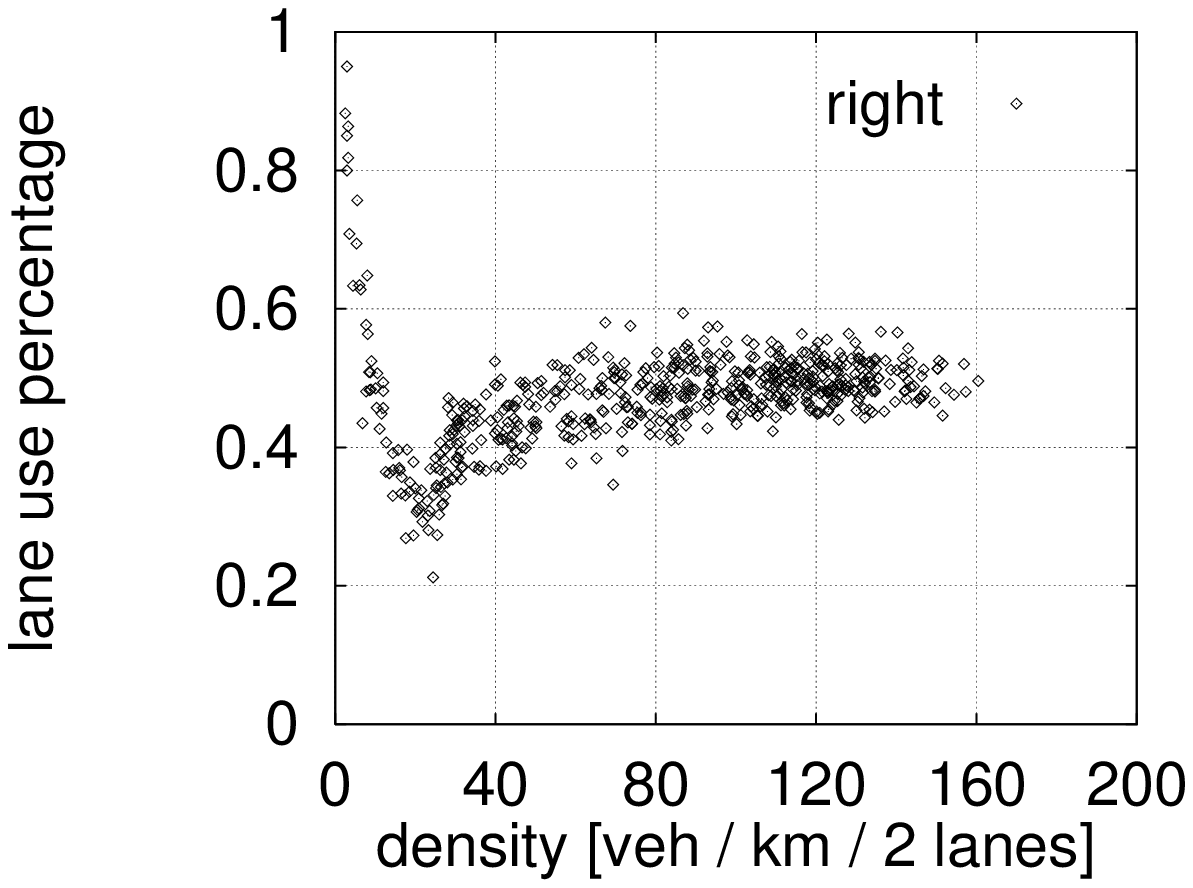}}

\caption{\label{slack-and-sym}%
Plots when slack is used and symmetry at low velocities included.
{\em Top:} Flow vs.\ density.  {\em Bottom:} Lane usage vs.\ density.
}
\end{figure}
In order to be able to tune the onset density as well as the amount of
lane inversion the second parameter ``slack'' has been introduced in
addition to the look-ahead.  This, however, has the side effect that
traffic never reverts to an equal lane usage, even at very high
densities, similar to what we obtained with the gap-rules above.  In
order to improve this, we make the rule-set symmetric at zero speed.
In technical terms, this means that a vehicle at speed zero only
checks if the speed on the other lane is higher than on its own lane,
and if so, attempts to change lanes (restricted by the security
criterion).  Other solutions are possible to achieve this (see, e.g.,
Ref.~\cite{Wagner:etc:2lane}; or one could attempt to make the
look-ahead distance velocity-dependent, e.g.~$\propto v$).
Fig.~\ref{slack-and-sym} shows that our approach indeed works, i.e.\
the lane usage at high densities now goes indeed to approximately 50\%
for each lane.

\subsection{Slow vehicles}
\begin{figure}
\centerline{\epsfxsize0.49\hsize\epsfbox{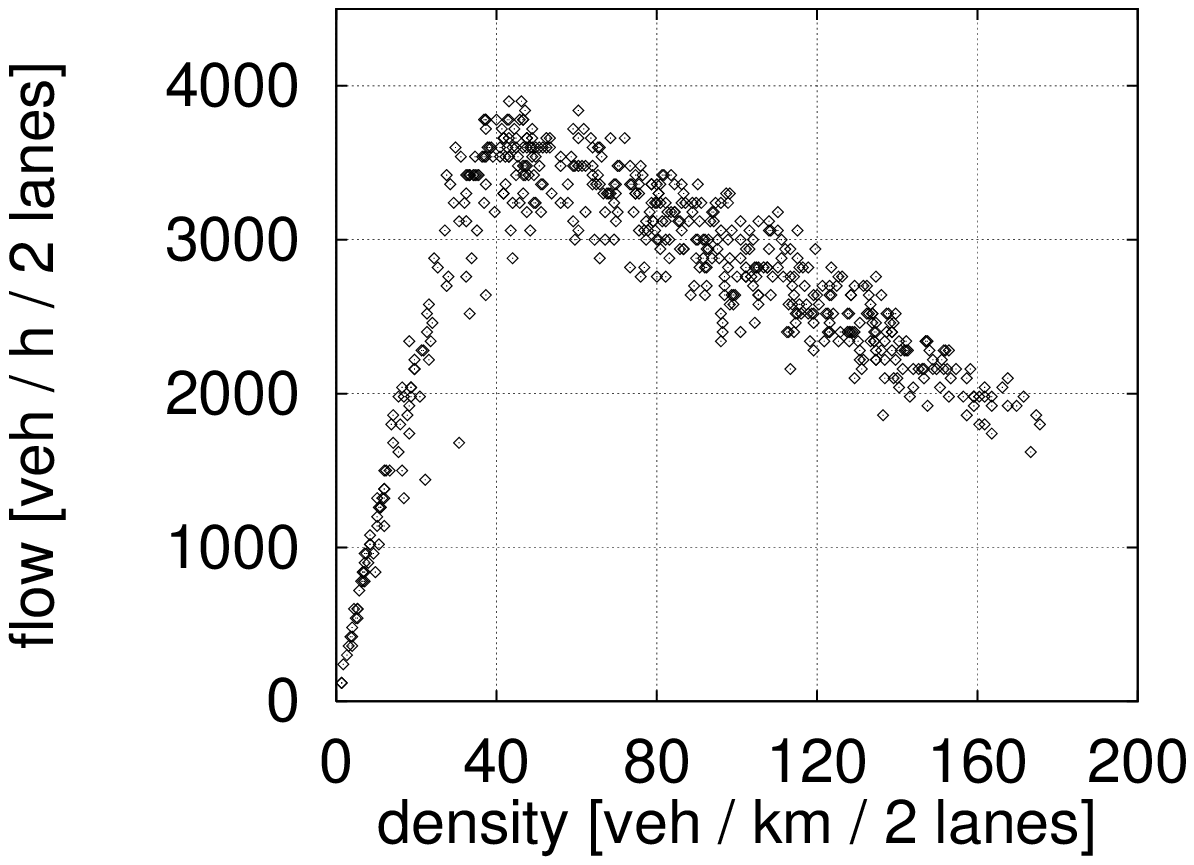}} 
\centerline{\epsfxsize0.49\hsize\epsfbox{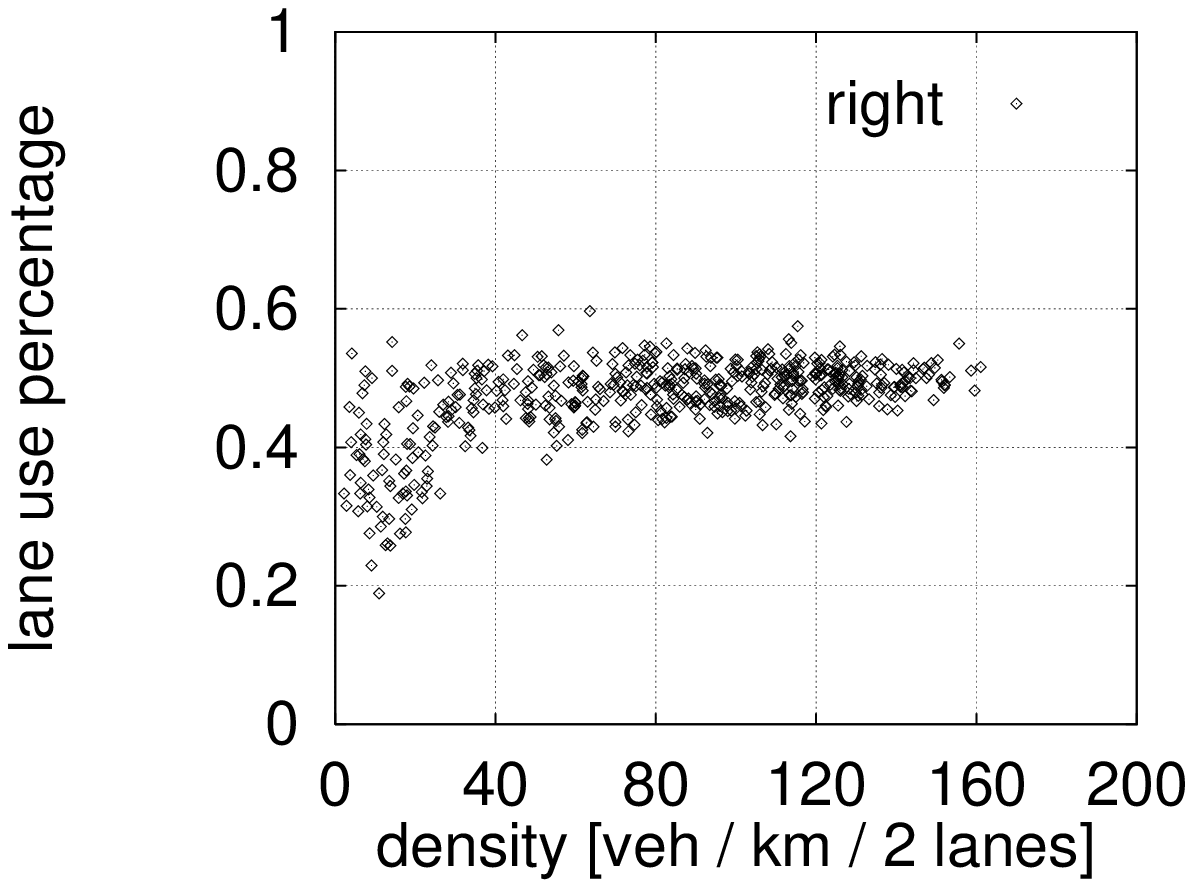}} 

\caption{\label{slow}%
Plots when slow vehicles included.  {\em Top:} Flow vs.\ density.
{\em Bottom:} Lane usage vs.\ density.
}
\end{figure}
Wiedemann's data includes 10~percent trucks.  We model the effect of
trucks by giving 10~percent of the vehicles a lower maximum
velocity~\cite{Rickert:diplom,Latour,Chowdhury:etc:2lane}.  Note
that this only models the lower speed limit which is in effect for
trucks in most European countries, but not the lower acceleration
capabilities.  The result for the flow-density curve and for the lane
usage is shown in Fig.~\ref{slow}.  The main difference to before is
that the maximum flow is shifted towards higher densities, and there
are more fluctuations in that region~\cite{Rickert:diplom}.

\subsection{Combination of all extensions}
\begin{figure}

\centerline{\hfill\epsfxsize0.49\hsize\epsfbox{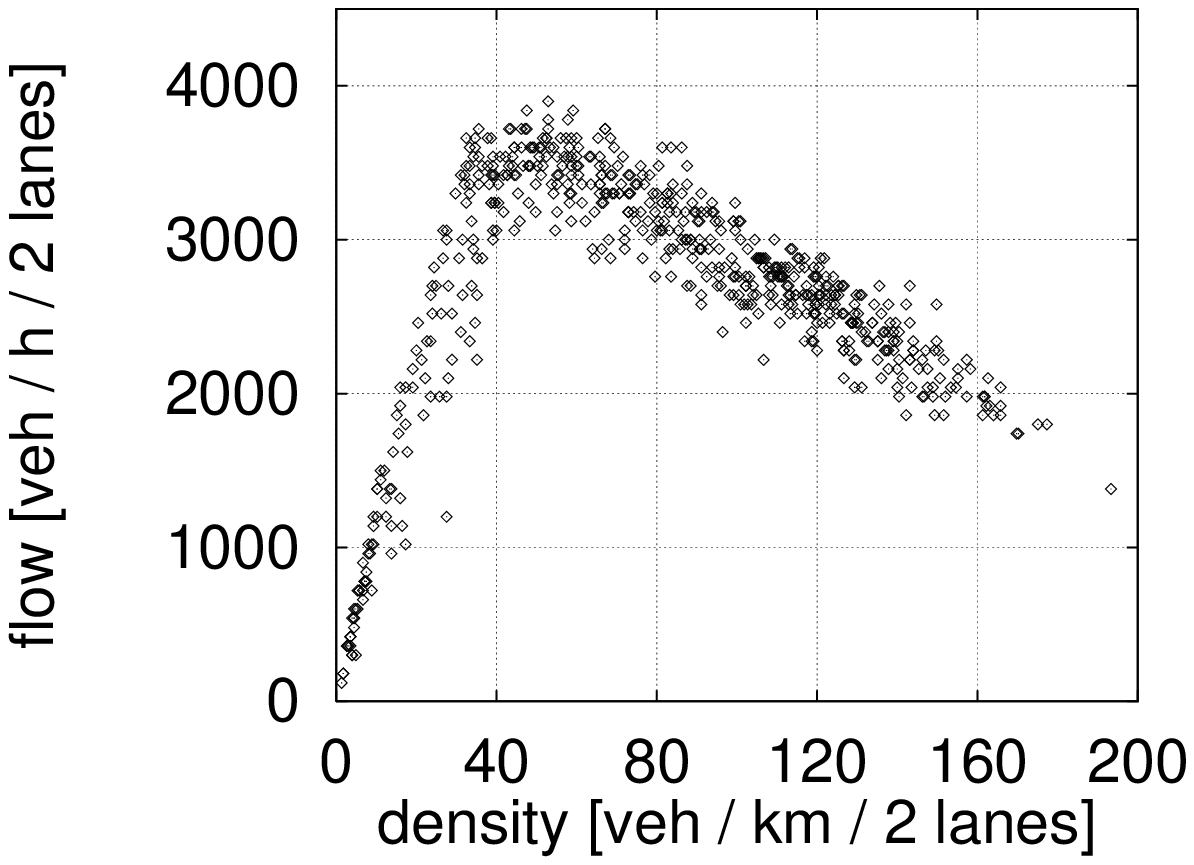}}
\centerline{\epsfxsize0.49\hsize\epsfbox{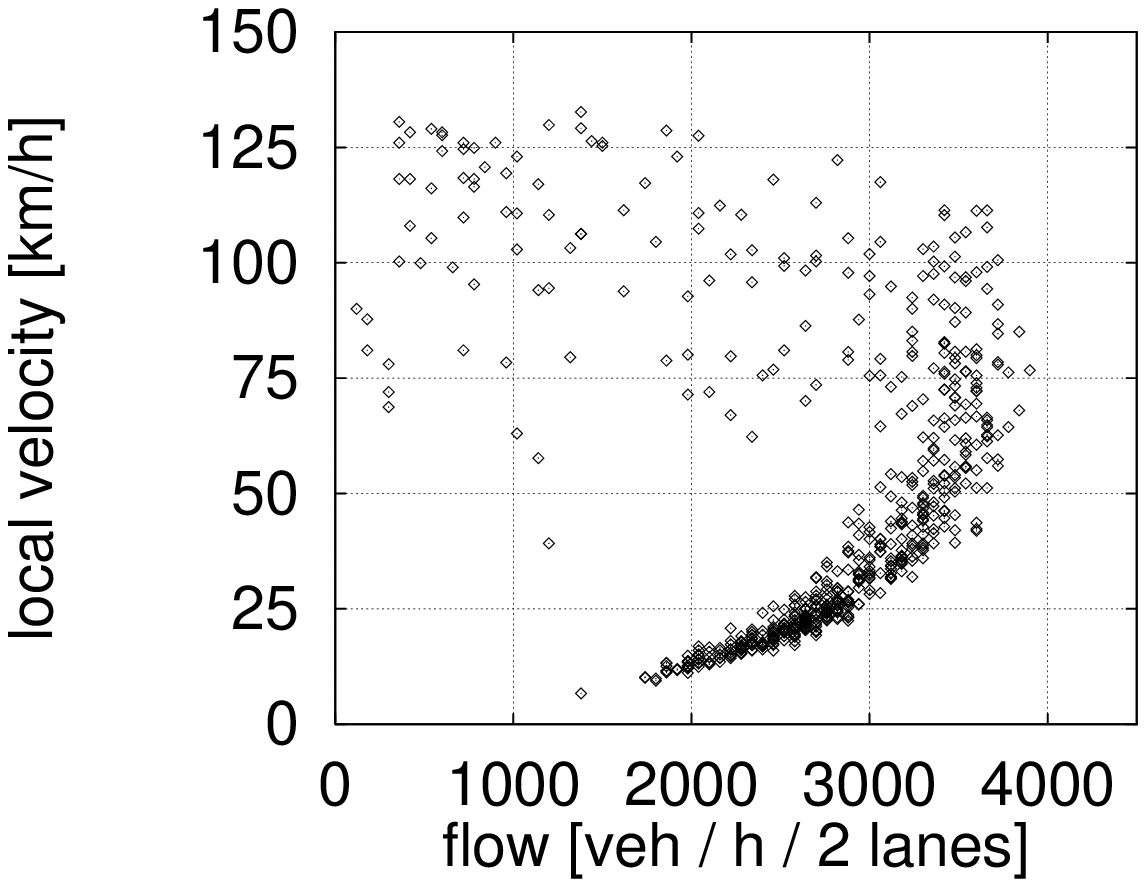} \hfill \epsfxsize0.49\hsize\epsfbox{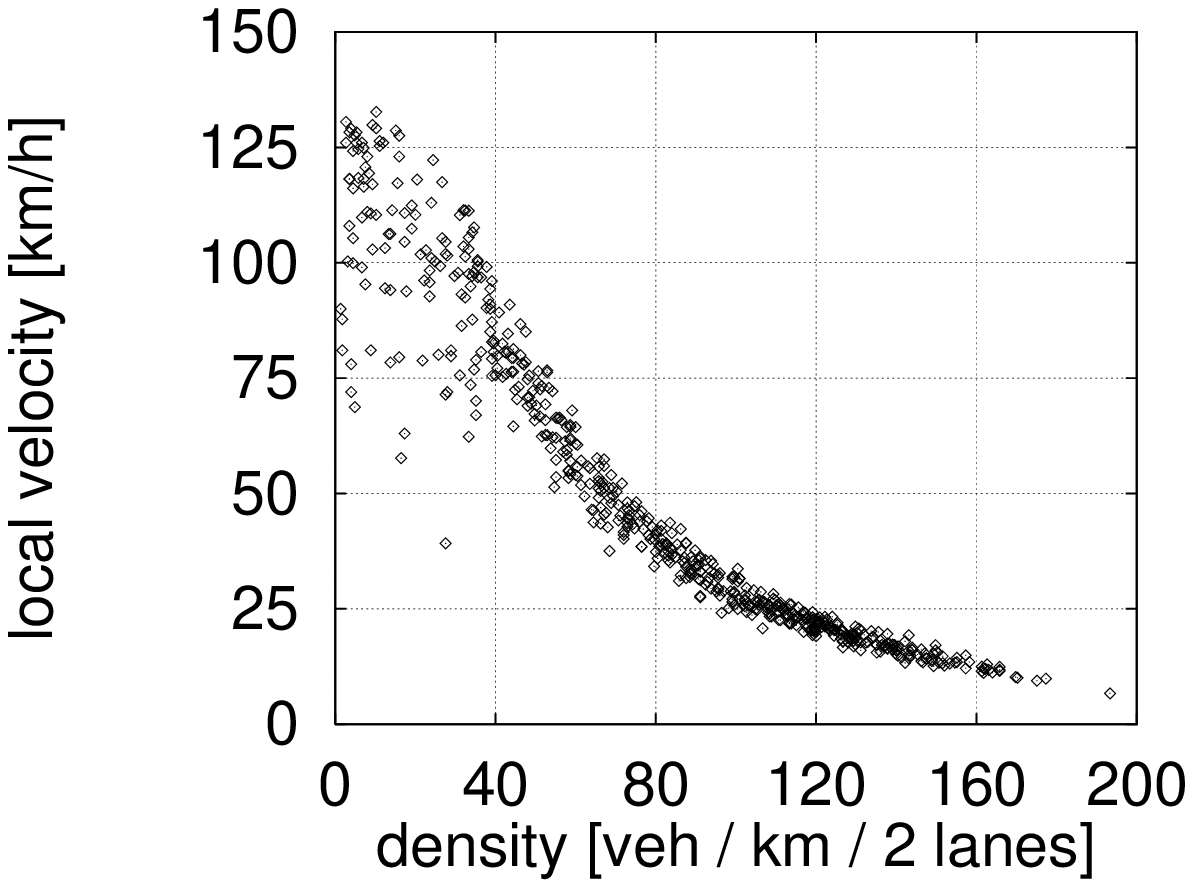}}
\centerline{\epsfxsize0.49\hsize\epsfbox{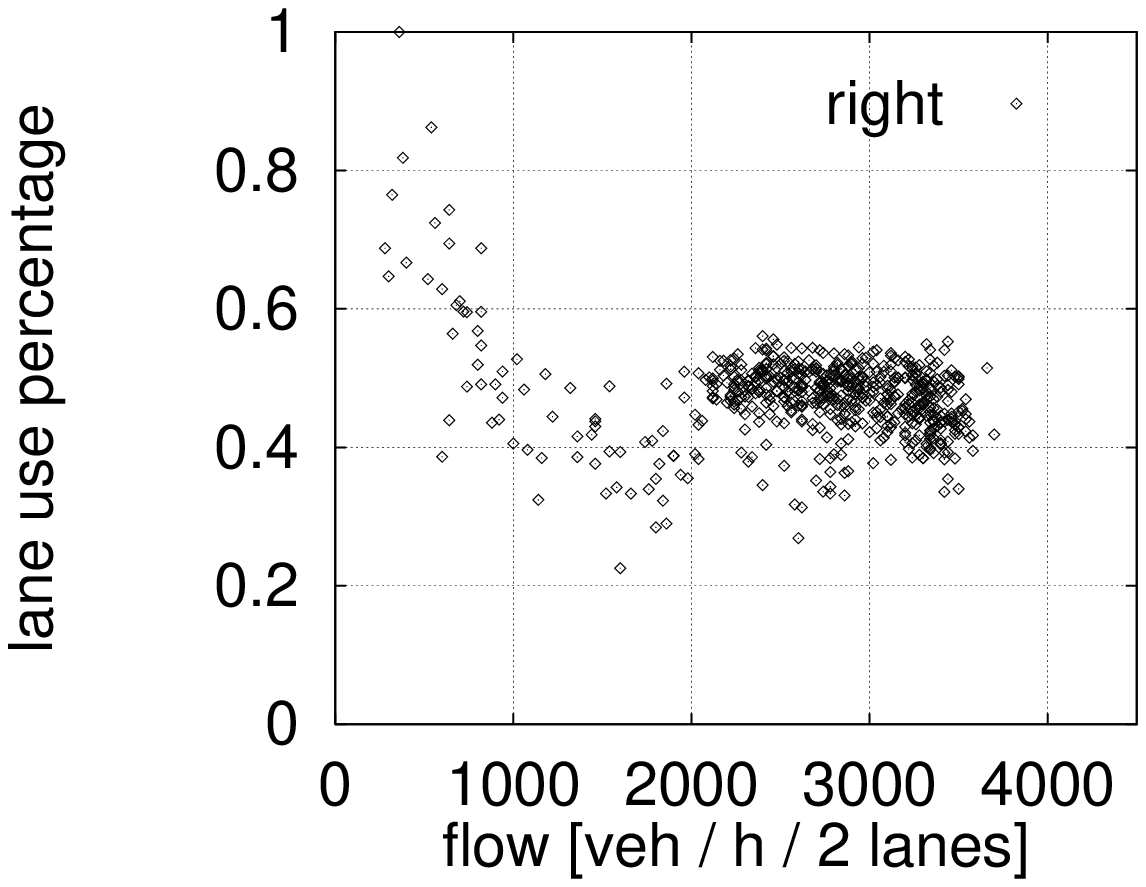} \hfill \epsfxsize0.49\hsize\epsfbox{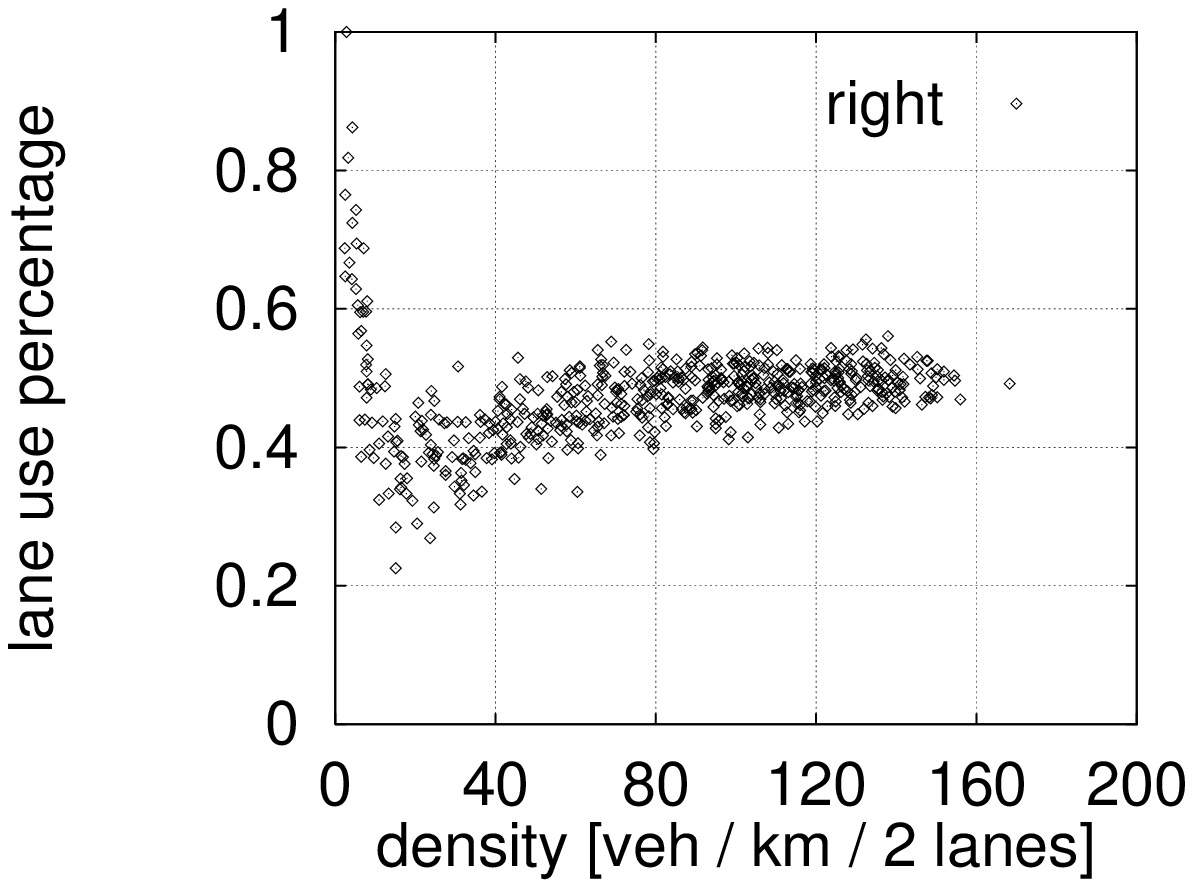}}

\caption{\label{all}%
All three extensions of the basic rule--set (trucks; symmetry at high
density; slack) are included.  Compare this figure to
Fig.~\protect\ref{Wiedemann}.
}
\end{figure}
Last, we show simulation results where all the above improvements
(trucks; symmetry at high densities; slack) are used simultaneously
(Fig.~\ref{all}).  Indeed, the results are now close to reality (cf.\
Fig.~\ref{Wiedemann}).

\section{The flow breakdown mechanism near maximum flow}
One of the questions behind this research was to investigate if, in highly
asymmetric two-lane systems, flow breakdown is indeed triggered by a single
lane flow breakdown on the left lane.  In order to address this
question, we will, in the following, study space-time plots of the
respective traffic dynamics as well as fundamental diagrams by lane.
Since it turns out that traffic without slow vehicles is fundamentally
different from traffic with slow vehicles, we will treat the two
situations separately.

\subsection{Maximum flow without slow vehicles}
\begin{figure}

\centerline{\epsfxsize0.65\hsize\epsfbox{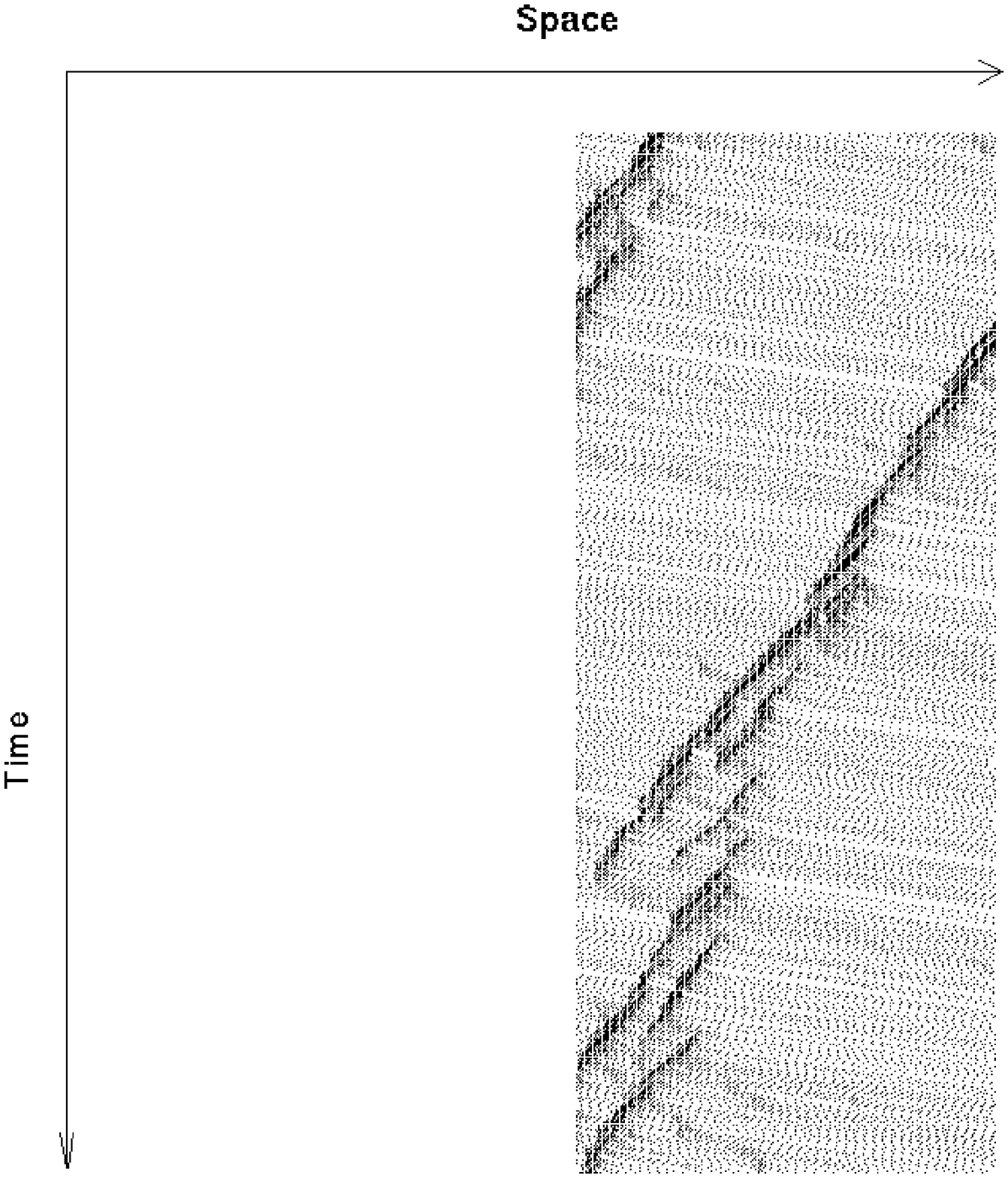}}

\caption{\label{1-lane:wo:slow:pixel}
Space-time plot of one-lane traffic without slow vehicles.
}
\end{figure}
\begin{figure}

\centerline{\epsfxsize0.65\hsize\epsfbox{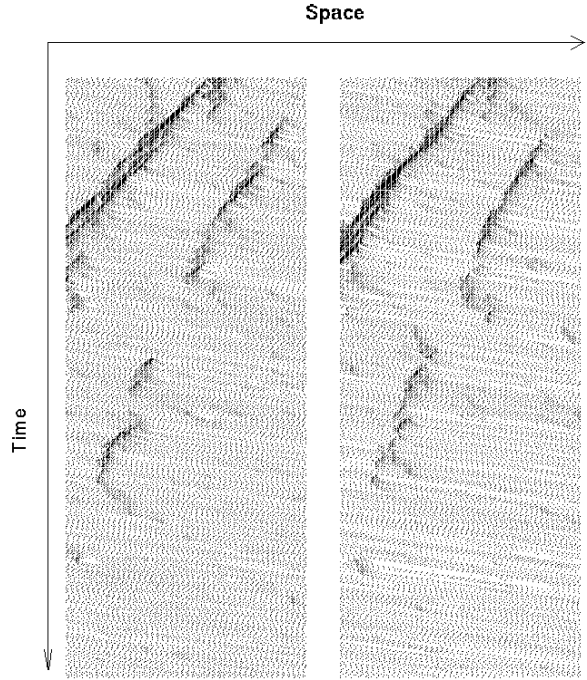}}

\caption{\label{basic:pixel}%
Space-time plot of two-lane traffic with the ``basic'' lane changing
rules without slow vehicles.  \emph{Left:} left lane. \emph{Right:}
right lane.
}
\end{figure}
\begin{figure}
\centerline{\epsfxsize0.49\hsize\epsfbox{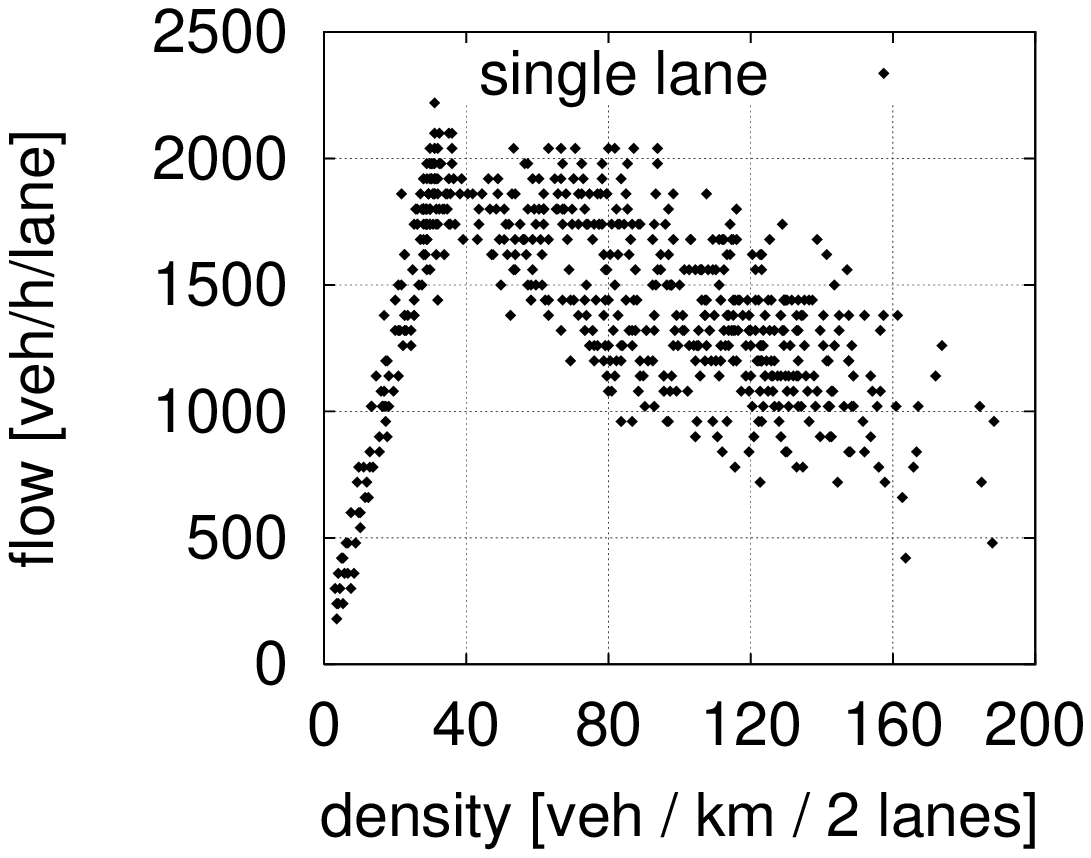}}
\vspace{-2mm}
\centerline{\epsfxsize0.49\hsize\epsfbox{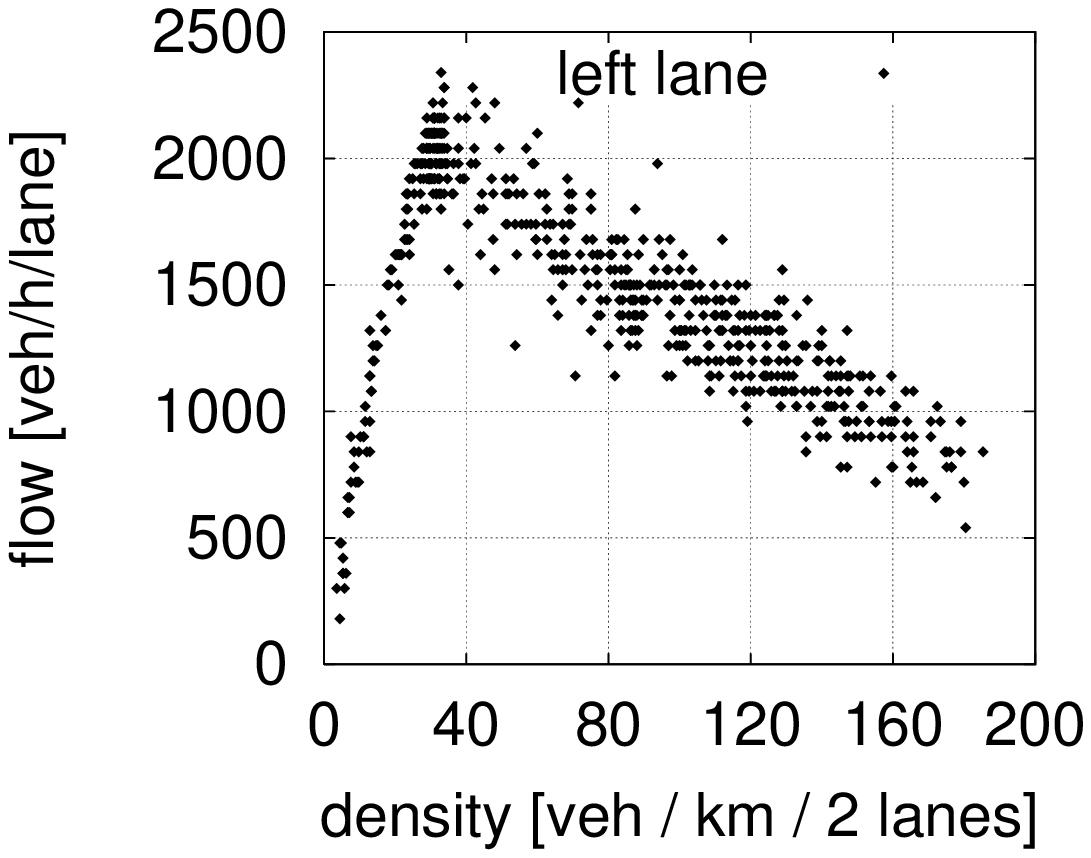}}
\vspace{-2mm}
\centerline{\epsfxsize0.49\hsize\epsfbox{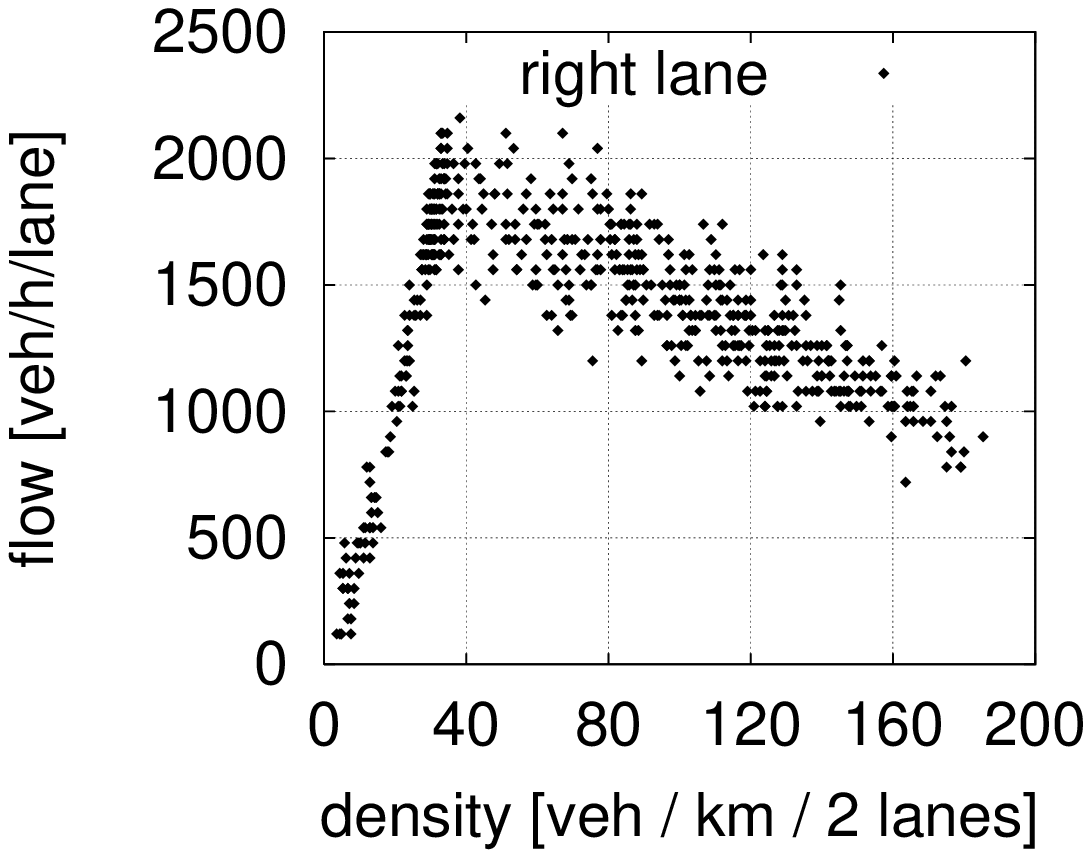}}
\caption{\label{1lane:and:basic:fdiags}%
{\em (a)} Fundamental diagram for single-lane rules.  {\em (b)}
Fundamental diagram for left lane of basic velocity two-lane rules,
i.e.\ plotting flow on the left lane vs.\ density on both lanes for
1-minute averages.  {\em (c)} Fundamental diagram for right lane of
basic velocity two-lane rules.
}
\end{figure}
Figs.~\ref{1-lane:wo:slow:pixel} and~\ref{basic:pixel} compare
space-time plots from a one-lane situation with the two-lane situation
using the ``basic'' velocity-based lane changing rules, in both cases
approximately at maximum flow.  Not much difference in the dynamics is
detectable except that maybe the 2-lane plot shows more small
fluctuations instead of fully developed jams.  This is confirmed by
the single-lane fundamental diagrams for the systems
(Fig.~\ref{1lane:and:basic:fdiags}): The fundamental diagram for the
left lane of basic velocity-based lane changing rules looks very
similar to the corresponding 1-lane diagram, and also the right lane
does not look much different.  Also, the density inversion has
reverted to 50:50 at maximum flow (Fig.~\ref{basic:v}).

Thus, the approach to maximum flow via increasing density is better
described in the way that the left lane reaches maximum flow earlier
than the right lane, and from then on all additional density is
squeezed into the right lane.  Only when the combined density of both
lanes is above the maximum flow density, flow break-down happens.
This argument gets confirmed by the observation that there are many
measurement points near maximum flow in all fundamental diagrams,
whereas at densities slightly higher than this significantly fewer
data points exist.  This should be compared to the situation which
includes slower vehicles, which will be explained next.

\subsection{Maximum flow with slow vehicles}
\begin{figure}

\centerline{\epsfxsize0.65\hsize\epsfbox{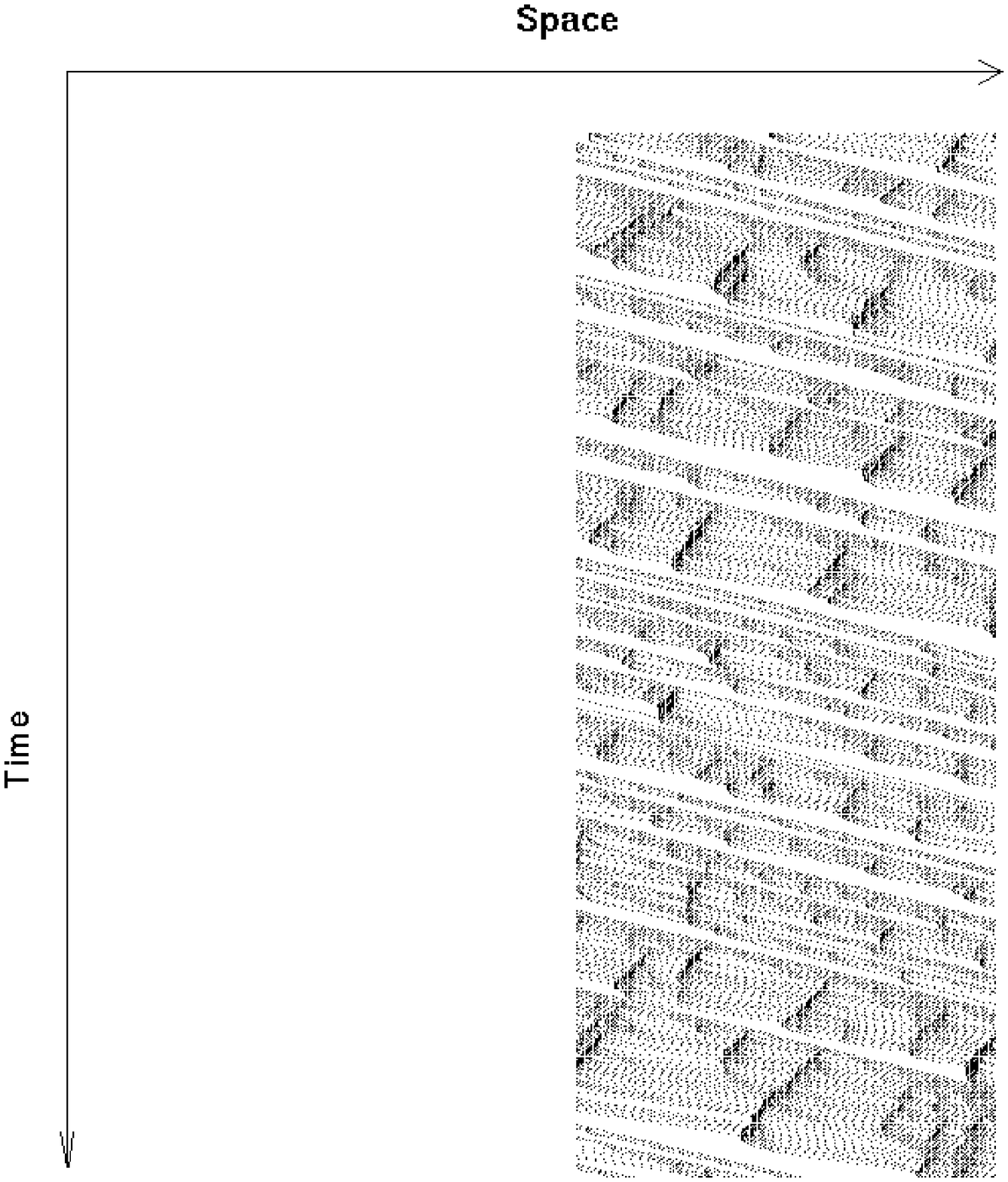}}


\caption{\label{1-lane-w-slow:pixel}%
Space-time plot of one-lane traffic near maximum flow including 10\%
slow vehicles.}
\end{figure}
\begin{figure}

\centerline{\epsfxsize0.65\hsize\epsfbox{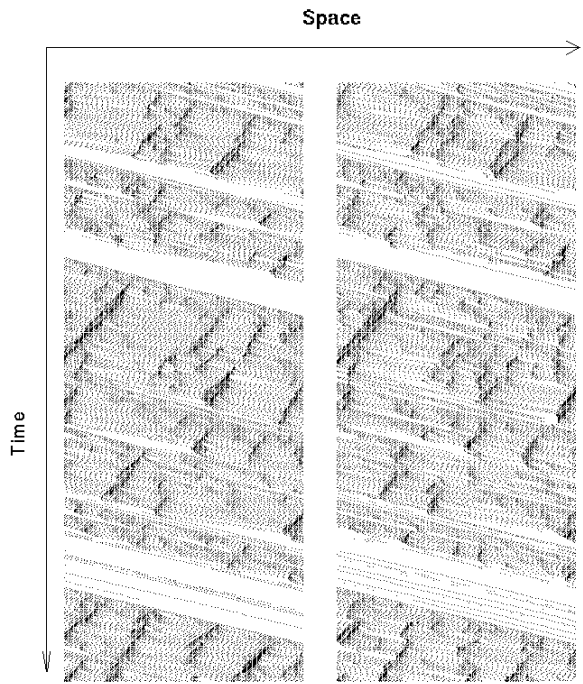}}

\caption{\label{basic-w-slow:pixel}%
Space-time plot of two-lane traffic near maximum flow including 10\%
slow vehicles using the ``basic'' velocity-based lane changing rules
of this paper.  \emph{Left:} left lane. \emph{Right:} right lane.}
\end{figure}
\begin{figure}

\centerline{\epsfxsize0.65\hsize\epsfbox{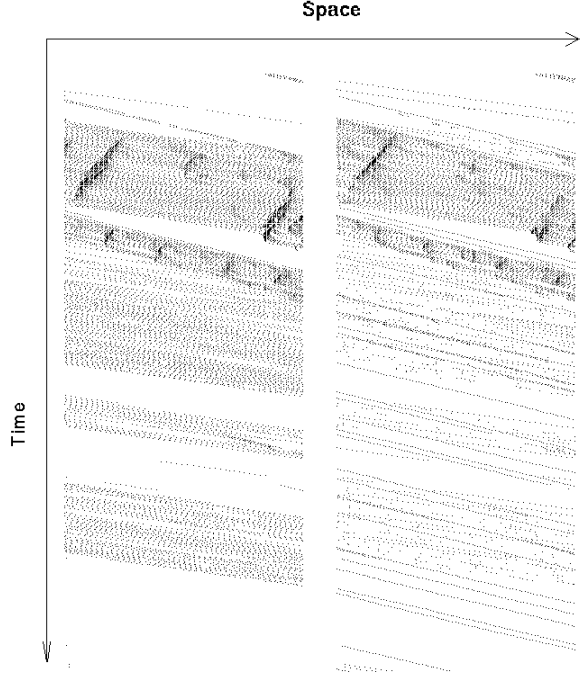}}

\caption{\label{basic-w-slow:pixel:lower-density}%
Space-time plot of two-lane traffic at about half the density of
maximum flow, including 10\% slow vehicles, using the ``basic'' lane
changing rules of this paper.  Same as
Fig.~\protect\ref{basic-w-slow:pixel}, except for the lower density.
\emph{Left:} left lane. \emph{Right:} right lane.}  
\end{figure}
\begin{figure}

\centerline{\epsfxsize0.65\hsize\epsfbox{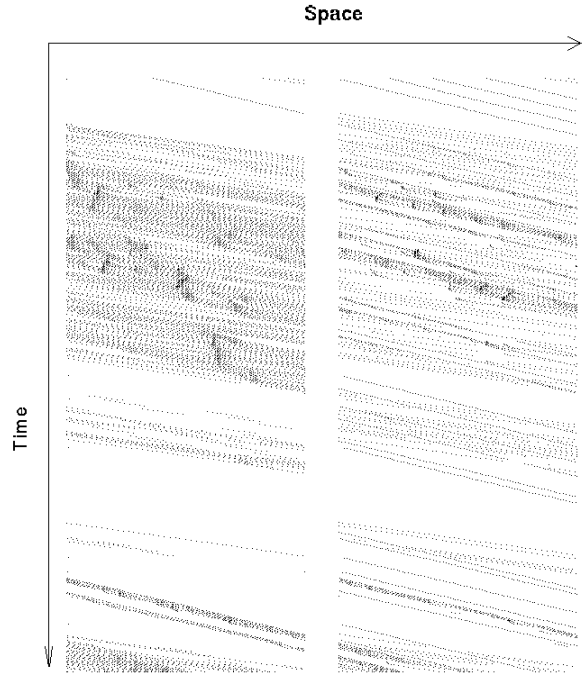}}

\caption{\label{all-w-slow:pixel:lower-density}%
Space-time plot of two-lane traffic at about half the density of
maximum flow, including 10\% slow vehicles, using the lane changing
rules with slack and symmetrization.  \emph{Left:} left
lane. \emph{Right:} right lane.  Trajectories of fast vehicles (less
steep slope) on the right lane which seem to go ``through'' the slow
vehicles (steep slope) are actually interrupted and go to the left
lane for short times.}
\end{figure}
\begin{figure}
\centerline{\epsfxsize0.49\hsize\epsfbox{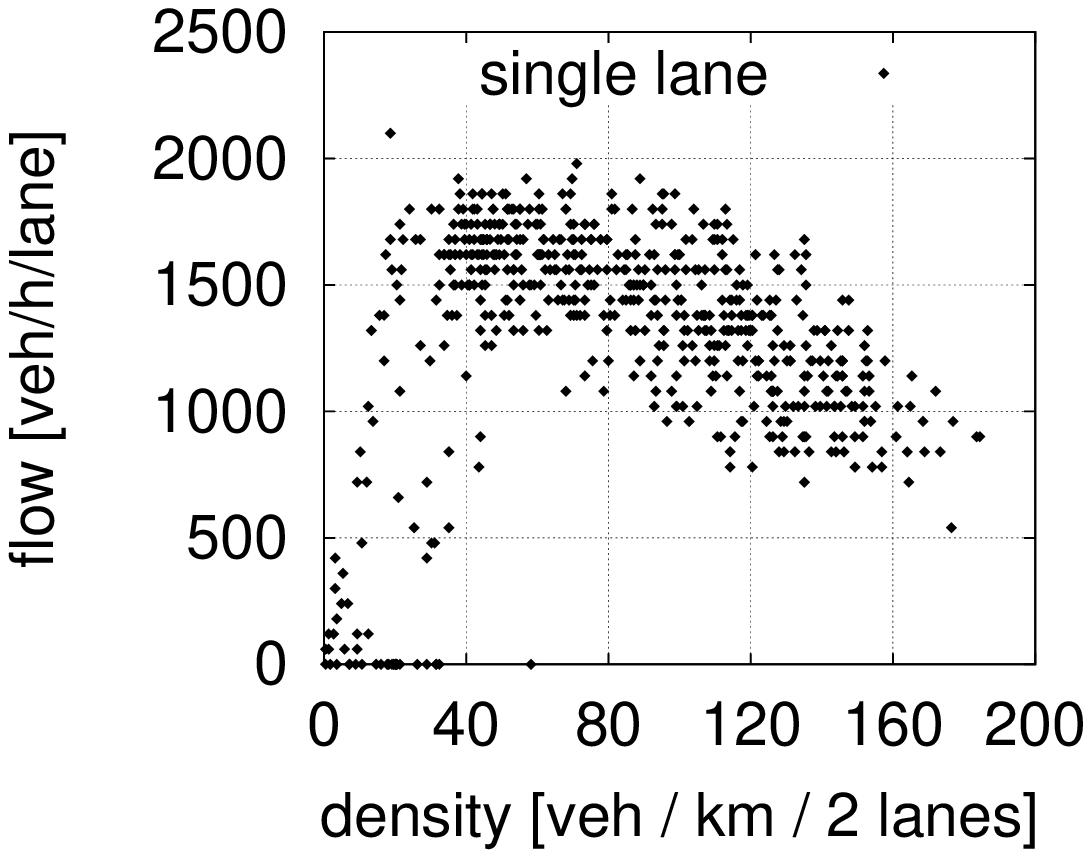}}
\vspace{-3mm}
\centerline{\epsfxsize0.49\hsize\epsfbox{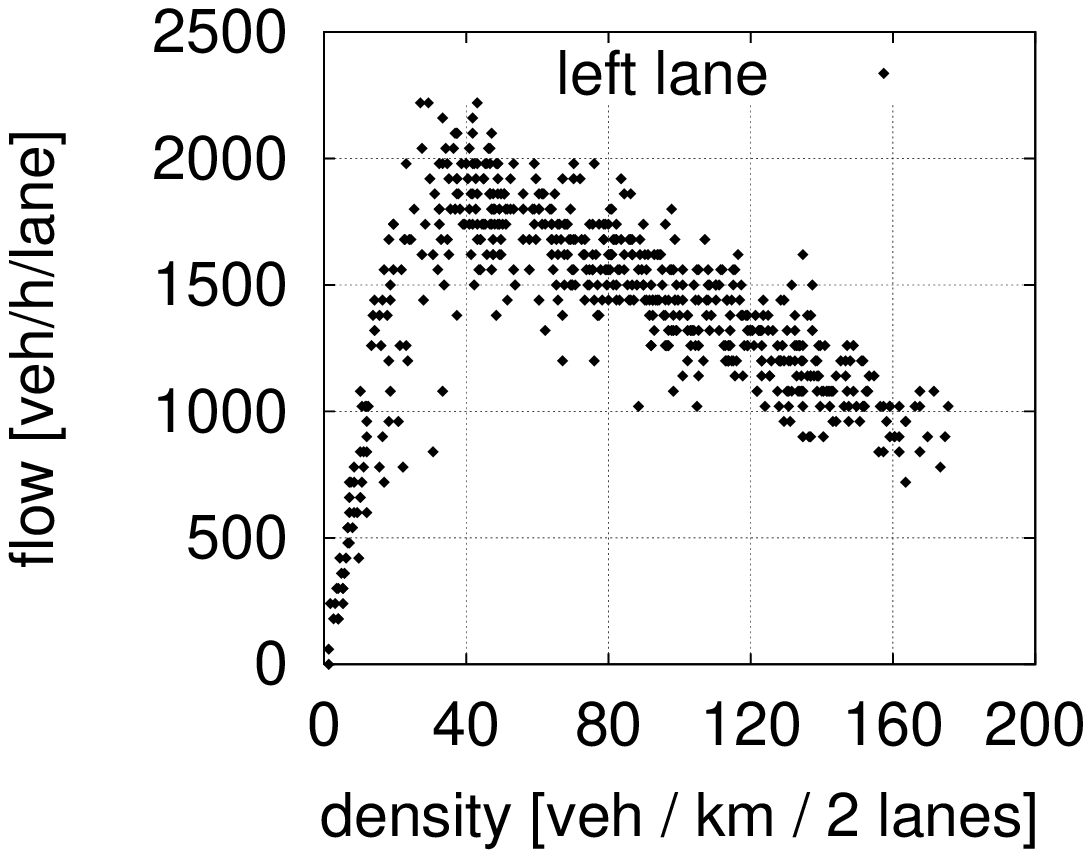}}
\vspace{-3mm}
\centerline{\epsfxsize0.49\hsize\epsfbox{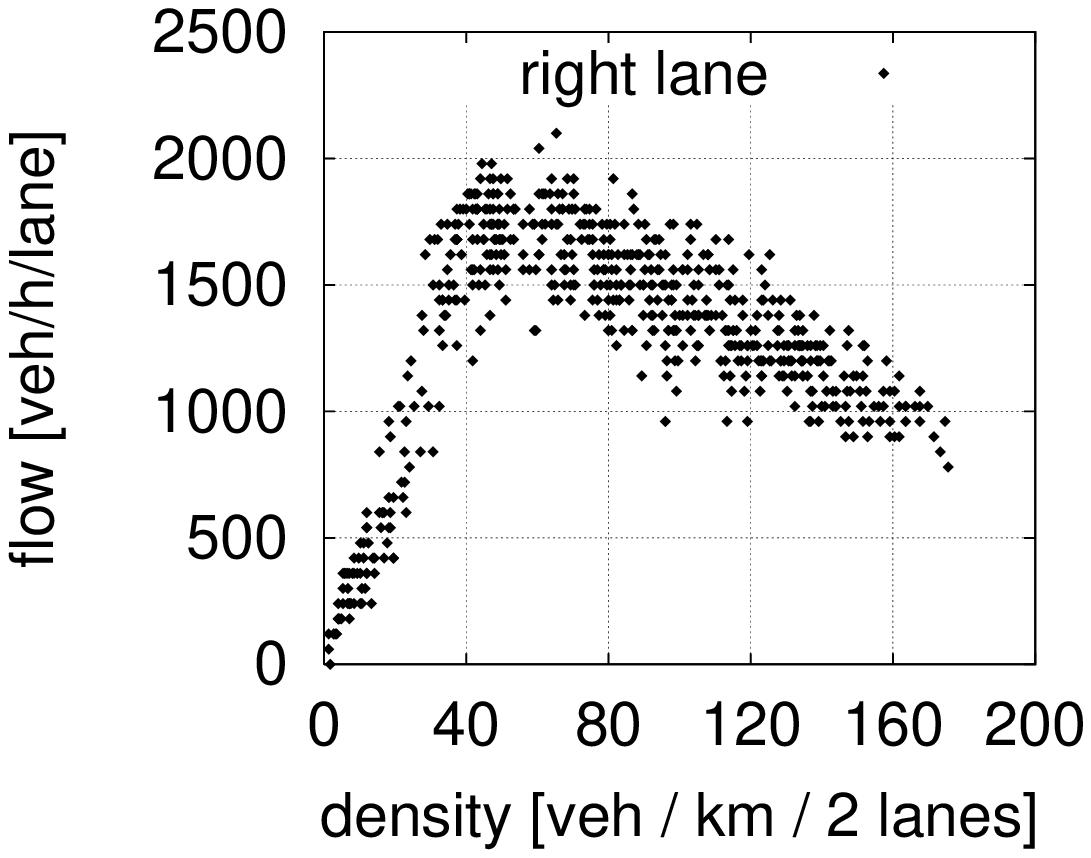}}
\caption{\label{w-slow:fdiags}%
Simulation results for traffic including 10\% slow vehicles: {\em (a)}
Fundamental diagram for single-lane rules.  {\em (b)} Fundamental
diagram for left lane of basic velocity-based two-lane rules, i.e.\
plotting flow on the left lane vs.\ density on the left lane for
1-minute averages.  {\em (c)} Fundamental diagram for right lane of
basic velocity two-lane rules.}
\end{figure}
The situation when slow vehicles are present is markedly different.
The 2-lane situation with slow vehicles
(Fig.~\ref{basic-w-slow:pixel}) looks more like the 1-lane situation
with slow vehicles~(Fig.~\ref{1-lane-w-slow:pixel}) than like the
2-lane situation without slow vehicles (Fig.~\ref{basic:pixel}).  That
means: The presence of slow vehicles has a stronger influence on the
dynamics than the difference between 1-lane and 2-lane traffic.  The
dominating feature is that fast vehicles jam up behind slow vehicles
and get involved in start--stop dynamics which gets worse with
increasing distance from the leading slow vehicle.  In the 2-lane
situation, these ``plugs'' are caused by two slow vehicles side by
side; a situation which is empirically known to happen regularly.

For the ``basic'' lane changing rules, the queues behind the ``plugs''
have similar length on both lanes, both near the density of maximum
flow (Fig.~\ref{basic-w-slow:pixel}) and at lower densities
(Fig.~\ref{basic-w-slow:pixel:lower-density}).  In contrast, when
using the lane changing rules with slack and symmetrization, then in
the same situation, there are more vehicles behind the truck on the
left than there are behind the truck on the right
(Fig.~\ref{all-w-slow:pixel:lower-density}).  Experience seems to
indicate that the more complicated rule-set is the more realistic one
here.

The lane-based fundamental diagrams (Fig.~\ref{w-slow:fdiags}) confirm
the observation that slow vehicles change the dynamics.  The marked
peak and the accumulation of data points near maximum flow are both
gone; maximum flow is found over a wider density range than before.
The flow on the left lane generally reaches higher values both than
flow on the right lane, and than single-lane traffic flow.

Space-time plots (Figs.~\ref{basic-w-slow:pixel}
and~\ref{basic-w-slow:pixel:lower-density}) show why this is the case.
Traffic in this situation is composed of two regimes:\begin{itemize}

\item
``Plugs'' of slow vehicles side by side, and faster vehicles queued up
behind them.

\item
``Free flow'' regions, where the slow vehicles stay on the right and
the fast vehicles are mostly on the left.

\end{itemize}
At low density, there are mostly free flow regions and a couple of
``plugs'' with queues behind them.  With increasing density, the share
of the free flow regions {\em de\/}creases while the share of the
queueing regions {\em in\/}creases.  Eventually, the free flow regions
get absorbed by the queueing regions, a 2-lane variant of the
mechanism described in Ref.~\cite{Krug:Ferrari,Ktitarev:etc:gaps}.

{From} visual inspection, it is clear that up to that density
(approx.\ 40~veh/km/2\,lanes) the left lane carries a higher flow
since it only has fast cars in the free flow regions.  Above this
density, it is clear that now also the slow vehicles get slowed down
by the end of the queue ahead of them.

\section{Discussion}

(i) In spite of widespread efforts, many earlier models were {\em
not\/} able to reproduce the lane inversion.  Why is that so?  The
reason is that the lane inversion is a subtle spatial correlation
effect: ``I stay on the left if there is somebody ahead on the left.''
Indeed, some of the earlier models~\cite{Schuett,Latour} do not
contain this crucial rule.  Sparmann~\cite{Sparmann:2lane} contains it but
still does not reproduce the density inversion; so one would speculate
that the weight for this rule was not high enough.

(ii) Real-world traffic seems to be more stable in the laminar regime
than our simulated two-lane traffic.  This can be seen in the
``overshoot'' (hysteresis, see Ref.~\cite{Treiterer:hysteresis}) of
the low-density branch of the flow-density-plot which is more
pronounced in reality than in the results of this paper.  The
single-lane model~\cite{Nagel:Schreckenberg} looked more realistic
here.  Yet, recent research shows that the hysteresis effect is
actually related to the structure of the braking rules of the
single-lane velocity rules~\cite{Krauss:etc:metastable,Krauss:Kreta}.
More precisely: In models with more refined braking rules the laminar
traffic does not break down that easily because small
disturbances can be handled by small velocity adjustments.

In this context, it should be stressed that, as mentioned above, our
plots actually show three minute averages for the lane usage plots
whereas all other plots are generated from one minute averages.  The
reason for this is that one minute averages for lane usage had so much
variance that the overall structure was not visible.  Yet, in reality
one minute averages are sufficient also for this quantity.  This
indicates that our models have, for a given two-lane density, a higher
variation in the lane usage than reality has. --- Also, the plots of
velocity vs.\ flow indicate that the range of possible velocities for
a given flow is wider in the simulations than in reality, again
indicating that for a given regime, our model accepts a wider range of
dynamic solutions than reality.

(iii) The fact that we needed space-time plots for resolving many of
the dynamical questions indicates that the methodology of plotting
short time averages for density, flow, and velocity, has shortcomings.
The reason has been clearly pointed out in recent research~\cite{Kerner:Konh:large:amplitude,Nagel:flow,Bando:etc:95}: Traffic
operates in distinctively different dynamic regimes, two of them being
laminar traffic and jammed traffic.  Averaging across time means that
often this average will, say, contain some dynamics from the laminar
regime and some dynamics from the jammed regime, thus leading to a
data point at some intermediate density and flow.

In transportation science, it seems that this problem is empirically known
because people are using shorter and shorter time averages (1-min averages
instead of 5-min averages used a couple of years ago or 15-min averages
used ten or more years ago).  It seems that one should try vehicle based
quantities.  Plotting $v/\Delta x$ as a function of $1/\Delta x$, where
$\Delta x$ is the front-bumper to front-bumper distance between two
vehicles, is still a flow-density plot, but now individualized for
vehicles.  Instead of just plotting data point clouds, one would now have
to plot the full distribution (i.e.\ displaying the number of ``hits'' for
each flow-density value).

\section{Other two-lane models}

It is possible to review earlier lane changing models in the view of the
scheme presented in this paper.  In general, classifying some of the
earlier rules into our scheme is sometimes difficult, but usually possible.
For example, when one uses
\[
\strut gap_r < v_{max} \ .OR. \ gap_l < v_{max}
\]
as a reason to change to the left, then the negation of that,
including ``slack'' $\Delta$, would be the reason to change to the
right.  Let us also use a security criterion as follows:
\[
gap_- = v_{back} + 1
\]
(i.e.\ the distance to the car behind on the other lane should be
larger than its velocity) and
\[
gap_+ = \min[ \, gap + 1 , v_{max} ]
\]
(i.e.\ the distance to the car ahead on the target lane should be
larger than either (i) the distance to the car ahead on the current
lane, or (ii) the maximum velocity).  With the exception of the
addition of the second part of the Incentive Criterion to change left,
these are exactly the same rules as used in Ref.~\cite{Wagner:etc:2lane}.

Note, though, that this is not completely trivial.  For example, the
incentive to change left ``$gap_l > gap_r$'' of Ref.~\cite{Wagner:etc:2lane} is now in the security criterion.  Also, for
changes from left to right, the forward part of the security criterion
could be left out, at least for the values of $\Delta$ which have been
used.   Quite generally, it can happen that a rule can fit into our
logical scheme, but parts of the decision tree can never be reached so
that parts of the rule can be omitted without changing anything. 

Indeed, many asymmetric lane changing rules investigated in the
literature can be viewed through our characterization.
Table~\ref{reasons} contains many asymmetric lane changing rules from
the traffic cellular automaton literature.  The underlined parts have
been added to make the rules completely fit into our scheme, i.e.\ to
make the incentive to change to the right the logical negation
(sometimes including ``slack'') of the incentive to change to the
left.  It would be interesting to test if the neglected part of the
rules would be used often or not if they were actually implemented.

\section{Summary}

This paper classifies the multitude of possible lane changing rules
for freeway traffic.
The first part of this follows Sparmann~\cite{Sparmann:2lane}: One
can separate the rules into an ``incentive to change lanes'' and a
security criterion, which asks if there is enough space available on
the target lane.

The second part of this is the observation that in countries with a
default lane and a passing lane, the incentive to change right is just
the logical negation of the incentive to change left, with possibly
some slack (inertia).

The security criterion seems to be universal for all reasonable lane
changing rules: $[-gap_-, gap_+]$ has to be empty on the target
lane; the exact values of the parameters $gap_-$ and $gap_+$ do not
seem to matter too much as long as they are reasonably large.  We used
$gap_- = v_{max}$ and $gap_+ = v$.

For the Incentive Criterion we argue that its general structure for
highly asymmetric traffic has to be ``change to the left when either
on your lane or on the left lane somebody is obstructing you'', and
``change back when this is no longer true''.  Since this usually leads
to a generic density inversion at high densities, one has to add a
symmetrizing rule for high density traffic.  We simply used a
symmetric Incentive Criterion for vehicles with velocity zero.

Both velocity and gap based implementations of this give satisfying
results.

Further, we showed that most asymmetric lane changing models in the physics
literature fit into this scheme.


\section*{Acknowledgements}

We thank R.~Wiedemann for making Fig.~\ref{Wiedemann} available to us.
This work has been performed in part under the auspices of the U.S.\
Department of Energy at Los Alamos National Laboratory, operated by
the University of California for the U.S. Department of Energy under
contract W-7405-ENG-36, and at the HLRZ, Forschungszentrum J\"ulich.
DW and KN also thank the Deutsche Forschungsgemeinschaft for support.

\begin{appendix}

\section{Transformation of the Wagner's rules from
Ref.~[28]}

Finding a correspondance for the rules of Wagner in
Ref.~\cite{Wagner:julich} is not straightforward.  However, at closer
inspection, the rules turn out to be inconsistent for certain choices
of parameters.  The forward part of the Incentive Criterion is:
\[
R \to L: gap_r < v_{max} ~.AND.~ gap_l > gap_r
\]
\[
L \to R: gap_r > v+\Delta' ~.AND.~ gap_l > v+\Delta'
\]
Assume for example a case where $gap_r = 3$, $gap_l = 4$, $v=0$,
$v_{max} \ge 4$, and $\Delta' = 0$.  Then the vehicle does not want to
be in either lane.

This problem gets resolved for $\Delta' \ge v_{max}-1$; and indeed
$\Delta' \ge 6$ was used.

Now, if one assumes $\Delta' \ge v_{max}-1$, then one can simplify the
rule-set.  One can move the condition $gap_l > gap_r$ into the security
criterion $gap_+ \ge \min[gap+1,v_{max}]$, and the remaining incentives to
change lanes are:
\[
R \to L: gap_r < v_{max} ~\underline{.OR.~ gap_l < v_{max}}
\]
\[
L \to R: gap_r \ge v_{max} + \Delta(v) ~.AND.~ gap_l \ge v_{max} + \Delta(v) \ ,
\]
where, as in Table~\ref{reasons}, the underlined part is added to make
the rule fit into the scheme.  Note that in this interpretation, the
slack now is $\Delta(v) = \Delta' - v_{max} + v$, i.e.\ a function of
the velocity.

\end{appendix}

\begin{table}
\epsfxsize\textwidth\epsfbox{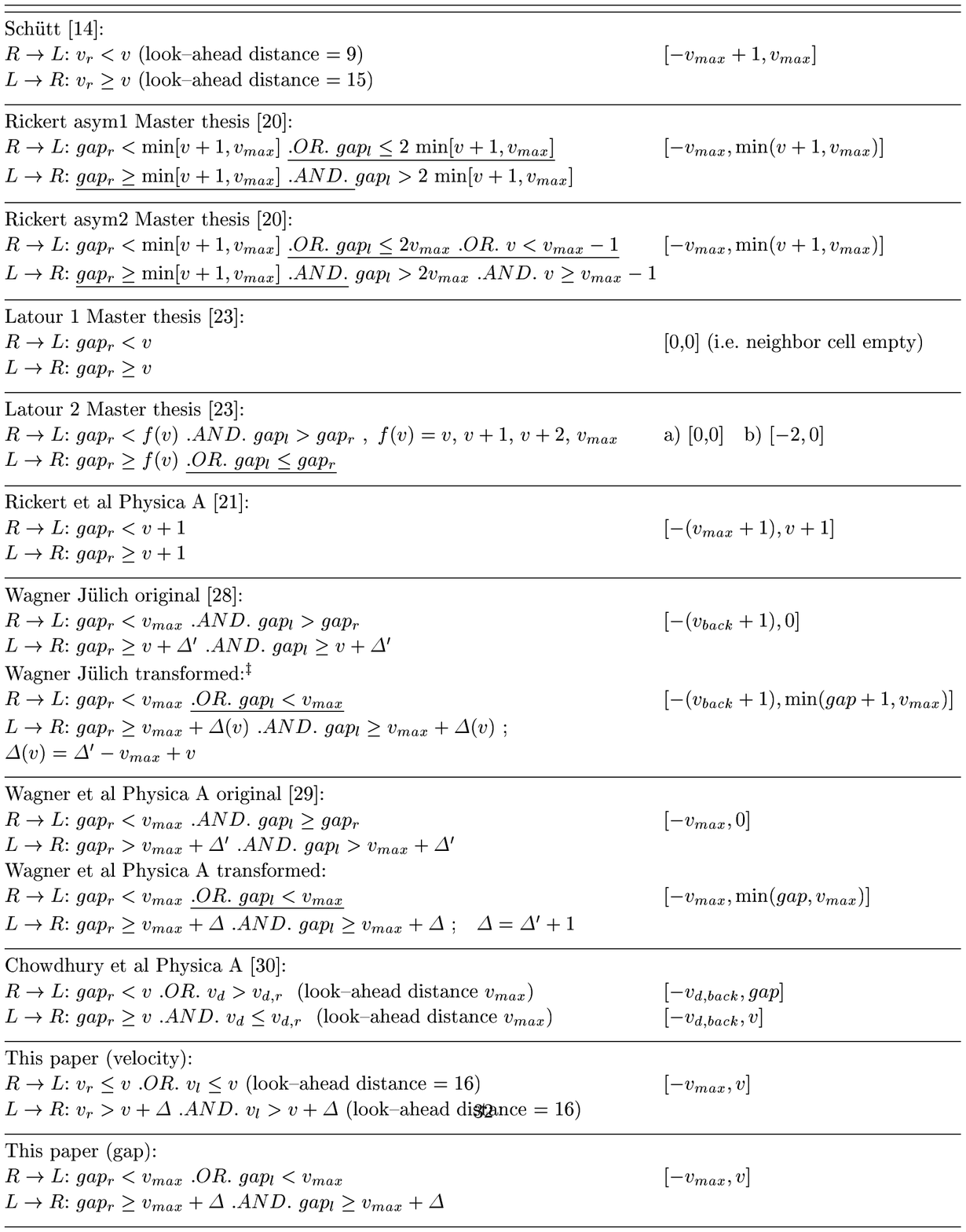}
\caption{\label{reasons}%
Lane changing rules in the literature.  The left column gives the
``incentives to change lane'' for the indicated lane change right to
left ($R \to L$) or left to right ($L \to R$).  The right colomn gives
the security criterion, i.e.\ the sites on the target lane that need
to be empty.  Underlined parts need to be added to make the incentive
to go right the logical negation of the incentive to go left.
``look--ahead distance'' is the distance to look ahead.  $v_{back}$ is
the velocity of the next vehicle behind on the target lane.  $v_d$ is
the desired speed (i.e.\ may be smaller than $v_{max}$ to denote a
slower vehicle class.  $v_{d,r}$ is the desired speed of the next
vehicle ahead on the right lane.  $v_{d,back}$ is the desired speed of
the next vehicle {\em behind\/} on the target lane.  ${}^\ddagger$\,See
appendix of this paper.
}
\end{table}

\end{document}